# On the Number of RF Chains and Phase Shifters, and Scheduling Design with Hybrid Analog-Digital Beamforming (Draft with more Results and Details)

Tadilo Endeshaw Bogale, *Member, IEEE*, Long Bao Le, *Senior Member, IEEE*

Afshin Haghighat and Luc Vandendorpe *Fellow, IEEE*

## Abstract

This paper considers hybrid beamforming (HB) for downlink multiuser massive multiple input multiple output (MIMO) systems with frequency selective channels. The proposed HB design employs sets of digitally controlled phase (fixed phase) paired phase shifters (PSs) and switches. For this system, first we determine the required number of radio frequency (RF) chains and PSs such that the proposed HB achieves the same performance as that of the digital beamforming (DB) which utilizes $N$ (number of transmitter antennas) RF chains. We show that the performance of the DB can be achieved with our HB just by utilizing $r_t$ RF chains and $2r_t(N - r_t + 1)$ PSs, where $r_t \leq N$ is the rank of the combined digital precoder matrices of all sub-carriers. Second, we provide a simple and novel approach to reduce the number of PSs with only a negligible performance degradation. Numerical results reveal that only $20 - 40$ PSs per RF chain are sufficient for practically relevant parameter settings. Finally, for the scenario where the deployed number of RF chains $(N_a)$ is less than $r_t$, we propose a simple user scheduling algorithm to select the best set of users in each sub-carrier. Simulation results validate theoretical expressions, and demonstrate the superiority of the proposed HB design over the existing HB designs in both flat fading and frequency selective channels.

**Index Terms**

Hybrid Analog-Digital Beamforming, Massive MIMO, Millimeter wave, Phase shifter, RF chain

## I. INTRODUCTION

Multiple input multiple output (MIMO) is one of the promising techniques for improving the spectral efficiency of wireless channels. To exploit the full potential of MIMO, one can leverage the conventional digital beamforming (DB). There are many DB design approaches developed in the past couple of decades. However, these approaches are designed mainly for few number of antennas (around 10) [1], [2]. It is shown that deployment of the massive number of antennas at the transmitter and/or receiver (massive MIMO) can significantly enhance the spectral and energy efficiency of microwave and millimeter wave (mmWave) wireless networks [3], [4].

In DB, one radio frequency (RF) chain is required for each antenna element at the base station (BS) and user equipment (UE) where an RF chain includes low-noise amplifier, down-converter, digital to analog converter (DAC), analog to digital converter (ADC) and so on [5], [6]. Thus, when the number of BS antennas $N$ is very large, the high cost and power consumptions of mixed signal components, like high-resolution ADCs and DACs, imply that dedication of a separate RF chain for each antenna is highly inefficient [3], [7]–[11]. For these reasons, beamforming design with limited number of RF chains has recently received significant attention. One approach of achieving this goal is to deploy beamforming at both the digital and analog domains, i.e., hybrid beamforming (HB). In the digital domain, beamforming can be realized at the baseband frequency whereas, in the analog domain, beamforming is implemented by using low cost phase shifters (PSs) at the RF frequency [12]–[14]. Different implementation aspects of the HB architecture can be found in [15] and [16].

In [17]–[19], a HB architecture is suggested for single user massive MIMO systems where matching pursuit (MP) algorithm is utilized [5]. In [20], a codebook based HB is proposed for wideband mmWave wireless networks. The codebooks are designed symmetrically for mitigating the possible beam shift due to the large differences of wave lengths at different sub-bands. In [21], a low complexity codebook based RF beamforming based on multi-level RF beamforming and level-adaptive antenna selection is considered. In [15], two types of sub-array HB architecture are considered; interleaved and localized sub-arrays. In the interleaved array, antenna elements in each sub-array scatter uniformly over the whole array whereas, in a localized array, antenna elements are adjacent to each other. In [22], HB designs utilizing interleaved and side-by-side sub-arrays (i.e., like in [15]) is proposed. This design is used for adaptive angle of arrival (AOA) estimation and beamforming by utilizing differential beam tracking and beam search algorithms.

In [23], hybrid precoding scheme for multiuser massive MIMO systems is considered. The paper employs the zero forcing (ZF) hybrid precoding where it is designed to maximize the sum rate of all users. In [8], a beam alignment technique using adaptive subspace sampling and hierarchical beam codebooks is proposed for mmWave cellular networks. A multi-beam selection precoding approach while exploiting the sparse characteristics of mmWave channels is employed in [24]. In [25], a beam domain reference signal design for downlink channel with HB architecture is proposed to maximize the gain in a certain direction around the main beam. In [26], a HB design using convex optimization is proposed for power minimization and maximization of the worst case signal to interference plus noise ratio (SINR) problems.

A HB design for 60 GHz application utilizing planar antenna arrays is considered to analyze the SINR of all user equipments (UEs) in [27]. The work of [28] considers a HB architecture and focuses on solving the minimization of the transmit power subject to the SINR constraints. Numerical algorithm based on semi definite programming (SDP) relaxation is proposed for examining the optimization problem. In [29], the sum rate maximization problem for the downlink massive MIMO systems is studied using the MP solution approach. In [30], a beam training (or beam steering) problem for the 60 GHz mmWave communications is formulated as a numerical optimization problem such that the received signal is maximized. The paper aims at identifying the optimal beam pair from a prescribed codebook with little overhead using numerical search approach. Recently, the joint optimization of the analog and digital beamforming matrices are considered in [31] to maximize the achievable rate with different practical constraints under the condition that each antenna can only be connected to a unique RF chain. And low complexity hybrid precoding scheme with sub-array architecture has been proposed to optimize the channel capacity in [32]. The considered scheme leverages the the idea of iterative successive interference cancellation (SIC) which allows parallelization.

**Motivation:** The DB "always" achieves optimal performance for any design criteria and channel. Therefore, any HB design "cannot" achieve better performance than that of DB [11], [15], [17]. The aforementioned HB designs examine their performance for specific design cri-teria(s) and/or channel model(s). For instance, the designs of [17], [23], [29] consider sum rate maximization problem and it is not clear how the performance of these HB designs behave for other design criteria. This limitation also arises in all the aforementioned HB designs. This is mainly due to the fact that these HB designs are problem dependent, and are not able to quantify the performances of their designs in terms of that of the DB for any design criteria even if PSs have sufficiently high (i.e., theoretically infinite resolution)[1]. From these explanations, we can understand that designing a HB ensuring the same performance as that of DB "for any design criteria and channel" has not been addressed in the aforementioned HB designs.

**Objectives:** In a multiuser setup, which is the focus of this paper, the problem of designing a HB while achieving the same performance as that of the DB can be addressed for the uplink or downlink channels. Since there is a duality between these channels, examining the above problem for either of the channels will be sufficient which motivates the current work to consider downlink channel [2]. For this channel, each user can have either single antenna or multiple antennas. In the current paper, we assume each user to be equipped with single antenna. Moreover, in a typical macro BS, transceivers are designed to operate in a considerable range of bandwidths. The large bandwidth and multipath nature of wireless channels in a cellular system motivates us to consider frequency selective channels. In addition, we assume that perfect channel state information is available at the BS. Under these setups and assumptions, the current paper considers the following problems:

**P1** For arbitrary transceiver optimization criteria and channel matrix, first we consider the design of a HB architecture while ensuring the same performance as that of the DB design. Specifically, we perform a study on the number of required RF chains and PSs under an intuitive design by assuming that PSs have infinite resolution ($\mathbf{P1}_A$). Then, we examine realizing the analog beamforming part of the HB designed in $\mathbf{P1}_A$ using practical constant phase PSs (CPPSs) and switches only with negligible performance loss ($\mathbf{P1}_B$)?

**P2** Since the considered system model is a multiuser wideband massive MIMO system, each of the users may use only part of the available spectrum. This motivates us to consider the user scheduling problem using the HB architecture designed in $\mathbf{P1}_B$.

[1]In the following, we will use the phrase "infinite resolution PS" to reflect that the resolution of the corresponding PS is sufficiently high such that the effect of phase error due to quantization can be neglected.

The existing works on hybrid beamforming design employ different architectures. In general, three architectures are commonly adopted; partially connected architecture as in [31]–[33], fully connected architecture as in [5], [17], [23], [29] and an architecture utilizing digitally controlled paired PSs as in [5], [6]. In the current paper, we have employed the modified version of hybrid architecture suggested in [6] as it is suitable to address the above two problems (the details of this modification is provided in Section IV). We would like to emphasize here that the current paper addresses **P1** and **P2** only "sub-optimally". Therefore, ensuring global optimality for these problems is still an open research topic. On the other hand, these problems are addressed for general channel matrix and carrier frequencies. Hence, the results of the current paper are valid both for microwave and mmWave massive MIMO applications. The current paper has the following main contributions:

1) We propose a HB design and determine the number of RF chains and PSs for multiuser and multicarrier massive MIMO systems such that the performances of the proposed HB design and the DB design are the same by assuming that PSs have infinite resolution. In particular, we show that the performance of the DB can be achieved with the proposed HB just by utilizing $N_p = 2r_t(N - r_t + 1)$ infinite resolution PSs and $r_t$ RF chains, where $N$ is the number of antennas at the BS and $r_t$ is the rank of the combined digital precoder matrices of all sub-carriers ($\mathbf{B}^d$). Then, we provide a novel and simple approach to realize the HB by employing $r_t$ RF chains and $N_{cp} \ll N_p$ CPPSs per each RF chain. Particularly, for finite analog precoding matrix precision level of $10^{-p}$ (detailed in Section IV-B), only $80p$ CPPSs are required per each RF chain. As will be clear in the simulation section, $10^{-0.25} - 10^{-0.5}$ accuracy is sufficient for practically relevant design problems. Thus, in practice only $20 - 40$ CPPSs are required per each RF chain.

2) From contribution (1), we can notice that the number of RF chains is still $r_t$ (i.e., rank of $\mathbf{B}^d$) which depends on many factors such as the number of UEs and their channel matrices, and precoder design criteria (e.g., sum rate, max min rate [2]). Due to this fact, the number of RF chains deployed at the BS ($N_a$) could be less than the rank of $\mathbf{B}^d$. In such a case, the DB cannot be realized with our HB. For these reasons, we examine **P2** and provide performance analysis by considering sum rate maximization problem while ensuring $\text{rank}(\mathbf{B}^d) \leq N_a$. Specifically, under the commonly used uniform linear array (ULA) channel model and ZF precoding, we have shown that the performance achieved by the HB and DB designs are the same when the angle of departure (AOD) of the channels of the scheduled users have some special structure which will be clear in *Lemma 2*.

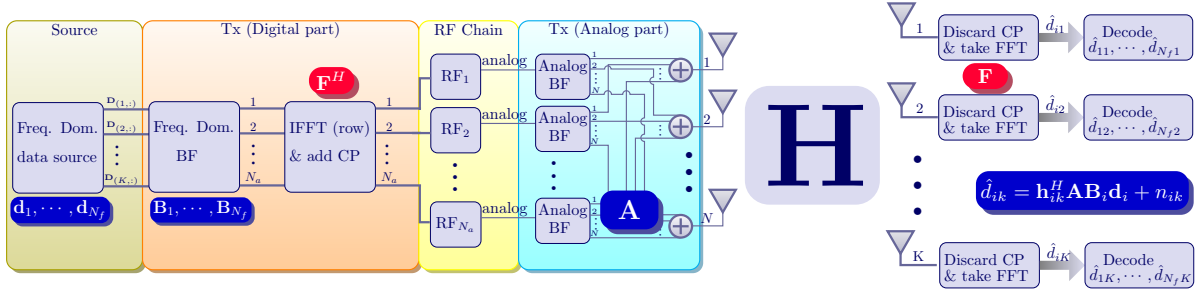

Fig. 1: System model of the proposed HB for multiuser and multicarrier systems.

3) We perform extensive numerical simulations to validate the theoretical results. We have also studied the effects of different parameters such as number of RF chains, BS antennas, PSs and total scheduled users on the performance of our design. Computer simulations also demonstrate that the proposed HB design achieves significantly better performance than those of the existing ones both for flat fading and frequency selective channels. Furthermore, the proposed design is convenient for practical realization of massive MIMO.

This paper is organized as follows. Section II discusses the HB system model. In Section III, a summary of Rayleigh fading and ULA channel models, and the conventional DB is provided. The proposed HB design is detailed in Section IV. In Sections V and VI, the proposed user scheduling and sub-carrier allocation algorithm, and its performance analysis is presented. Simulation results are provided in Section VII. Finally, we conclude the paper in Section VIII.

*Notations:* In this paper, upper/lower-case boldface letters denote matrices/column vectors. $\mathbf{X}_{(i,j)}$, $\mathbf{X}^T$, $\mathbf{X}^H$ and $\mathrm{E}(\mathbf{X})$ denote the $(i,j)$th element, transpose, conjugate transpose and expected value of $\mathbf{X}$, respectively. $\mathrm{diag}(.)$, $\mathrm{blkdiag}(.)$, $|.|$, $\lceil x \rceil$, $\mathbf{1}_N$, $\mathbf{I}$ and $\mathbb{C}^{N \times M}(\Re^{N \times M})$ denote diagonal, block diagonal, two norm, nearest integer greater than or equal to $x$, an $N$ sized vector of ones, appropriate size identity matrix and $N \times M$ complex (real) entries, respectively. The acronym s.t and i.i.d denote "subject to" and "independent and identically distributed", respectively.

## II. SYSTEM MODEL

This section discusses the proposed HB for a downlink multiuser and multicarrier massive MIMO system which is shown in Fig. 1. As we can see from this figure, the BS and each of the UEs are equipped with $N$ and 1 antenna, respectively. We employ block based multiuser orthogonal frequency domain multiple access (OFDMA) transmission where each block has $N_f$ sub-carriers. At each symbol period, the BS broadcasts $K$ symbols, where $K$ is the number of served UEs. Thus, in each OFDMA block, $K \times N_f$ symbols will be transmitted. For convenience, let us represent the transmitted symbols in each OFDMA block by $\mathbf{D} = [\mathbf{d}_1, \mathbf{d}_2, \cdots \mathbf{d}_{N_f}]$, where $\mathbf{d}_i = [d_{i1}, \cdots, d_{iK}]^T$ and $d_{ik}$ is the $k$th UE $i$th sub-carrier symbol. Since we have employed OFDMA transmission, $\mathbf{D}$ is the symbol matrix in frequency domain. The precoding and decoding operations of this frequency domain input data is explained as follows.

Let us define $\tilde{h}_{nk}^*(0), \cdots \tilde{h}_{nk}^*(L_p - 1)$ as the multipath channel coefficients between the $n$th BS antenna and $k$th UE, and $L_p$ is the number of multipath channel taps between the BS and all UEs. For this model, the $k$th UE received signal is given as [34]–[36]

$$\mathbf{r}_k^H = [\mathbf{h}_{1k}^H \mathbf{A} \mathbf{B}_1 \mathbf{d}_1, \cdots, \mathbf{h}_{N_f k}^H \mathbf{A} \mathbf{B}_{N_f} \mathbf{d}_{N_f}] \mathbf{F}^H + \tilde{\mathbf{n}}_k^H \quad (1)$$

where $\mathbf{B}_i \in \mathcal{C}^{N_a \times K}$ is the digital precoder matrix of the $i$th sub-carrier, $\mathbf{A} \in \mathcal{C}^{N \times N_a}$ is the analog precoder matrix, $\mathbf{F}^H$ is the inverse fast Fourier transform (FFT) matrix of size $N_f$, $\tilde{\mathbf{n}}_k^H \in \mathcal{C}^{1 \times N_f}$ is the noise vector at the $k$th UE and $\mathbf{h}_{ik}$ is the channel matrix of $k$th UE $i$th sub-carrier which is given as $\mathbf{h}_{ik} = [\mathbf{D}_{h1k}(i), \mathbf{D}_{h2k}(i), \cdots, \mathbf{D}_{hNk}(i)]^T$ with $\mathbf{D}_{hnk}^H = \mathrm{diag}(\lambda_{nk}(\{i\}_{i=0}^{N_f-1}))$ as a diagonal matrix of size $N_f$ and $\lambda_{nk}(i) = \sum_{s=0}^{L_p-1} \tilde{h}_{nk}^*(s) e^{-j\frac{2\pi i}{N_f} s}$. At the $k$th UE, the time domain signal will be transformed to frequency domain by employing FFT operation. It follows

$$\tilde{\mathbf{r}}_k^H = \mathbf{r}_k^H \mathbf{F} = [\mathbf{h}_{1k}^H \mathbf{A} \mathbf{B}_1 \mathbf{d}_1, \cdots, \mathbf{h}_{N_f k}^H \mathbf{A} \mathbf{B}_{N_f} \mathbf{d}_{N_f}] + \tilde{\mathbf{n}}_k^H \mathbf{F}. \quad (2)$$

The recovered signal of the $k$th UE's $i$th sub-carrier can now be expressed as

$$\hat{d}_{ik} = \mathbf{h}_{ik}^H \mathbf{A} \mathbf{B}_i \mathbf{d}_i + n_{ik}, \ \forall i, k \quad (3)$$

where $n_{ik} = \tilde{\mathbf{n}}_k^H \mathbf{f}_i$ is the $k$th UE $i$th sub-carrier noise sample which is assumed to be i.i.d zero mean circularly symmetric complex Gaussian (ZMCSCG) random variable with unit variance. The current paper assumes that $\mathbf{A}$ is realized with unity modulus PSs only as in [17], [23], [29]. Note that since the current paper assumes a single antenna UE, the UE's operation is the same as that of the conventional DB (i.e., HB is not required at the UE side).

## III. CHANNEL MODEL AND DIGITAL BEAMFORMING

For better exposition of the paper, this section summarizes the geometrical channel model and conventional DB.

### A. Channel Model

To model the $i$th sub-carrier channel between the BS and $k$th UE, we consider the most widely used geometric channel model with $L_s$ scatterers. Under this assumption, $\tilde{\mathbf{h}}_k(q) = [\tilde{h}_{1k}(q), \tilde{h}_{2k}(q), \cdots, \tilde{h}_{Nk}(q)]^T$ can be expressed as [8], [15], [17], [20], [29]

$$\tilde{\mathbf{h}}_k(q) = \sqrt{\frac{N}{L_s \rho_k}} \sum_{m=1}^{L_s} c_{km}(q) \tilde{\boldsymbol{\tau}}_k(\theta_{km}) = \boldsymbol{\tau}_k \mathbf{c}_k(q) \quad (4)$$

where $\rho_k$ is the distance dependent pathloss between the BS and $k$th UE, $c_{km}$ is the complex channel coefficient of the $k$th UE $m$th path with $\mathrm{E}\{|c_{km}|^2\} = 1$, $\theta_{km} \in [0,\ 2\pi]$ is the AOD, $\tilde{\boldsymbol{\tau}}_k(.)$ is the antenna array response vector of the $k$th UE, $\boldsymbol{\tau}_k = [\tilde{\boldsymbol{\tau}}_k(\theta_{k1}), \tilde{\boldsymbol{\tau}}_k(\theta_{k2}), \cdots, \tilde{\boldsymbol{\tau}}_k(\theta_{kL_s})]$ and $\mathbf{c}_k(q) = \sqrt{\frac{N}{L_s \rho_k}}[c_{k1}(q), c_{k2}(q), \cdots, c_{kL_s}(q)]^T$.

For performance analysis (discussed in Section VI), this paper adopts the most widely used Rayleigh fading and ULA channel models. The model (4) turns out to be Rayleigh fading channel when $L_s$ is very large, and (4) turns out to be ULA channel when $\tilde{\boldsymbol{\tau}}_k(.)$ is modeled as [8]

$$\tilde{\boldsymbol{\tau}}_k(\theta) = \frac{1}{\sqrt{N}}[1, e^{j\frac{2\pi}{\lambda}\tilde{d}\sin(\theta)}, e^{j2\frac{2\pi}{\lambda}\tilde{d}\sin(\theta)}, \cdots, e^{j(N-1)\frac{2\pi}{\lambda}d\sin(\theta)}]^T \quad (5)$$

where $j = \sqrt{-1}$, $\lambda$ is the transmission wave length and $\tilde{d}$ is the antenna spacing.

### B. Digital Beamforming

For better understanding of the proposed HB design, this subsection provides a brief summary on the structure of the DB matrix which is obtained by employing $N$ RF chains. Assume that we have employed DB approach to get the precoder matrices of all sub-carriers for an arbitrary design criteria. With the DB, the recovered data $\hat{d}_{ik}$ can be expressed as

$$\hat{d}_{ik}^d = \mathbf{h}_{ik}^H \mathbf{B}_i^d \mathbf{d}_i + n_{ik}, \ \forall i, k \quad (6)$$

where $\mathbf{B}_i^d$ is the digital precoder matrix of sub-carrier $i$. By taking the QR decomposition of the combined precoder matrix $\mathbf{B}^d = [\mathbf{B}_1^d, \mathbf{B}_2^d, \cdots, \mathbf{B}_{N_f}^d]$, one can get $\mathbf{B}^d = \mathbf{Q}^d \tilde{\mathbf{B}}^d$, where $\mathbf{Q}^d \in \mathcal{C}^{N \times r_t}$ is a unitary matrix which satisfies $(\mathbf{Q}^d)^H\mathbf{Q}^d = \mathbf{I}_{r_t}$, $\tilde{\mathbf{B}}^d \in \mathcal{C}^{r_t\times(K\times N_f)}$ is an upper triangular matrix and $r_t \leq N$ is the rank of the matrix $\mathbf{B}^d$. Hence, $\hat{d}_{ik}$ can be equivalently expressed as

$$\hat{d}_{ik}^d = \mathbf{h}_{ik}^H \mathbf{Q}^d \tilde{\mathbf{B}}_i^d \mathbf{d}_i + n_{ik}, \quad \forall i,k \tag{7}$$

where $\tilde{\mathbf{B}}_i^d$ is the sub-matrix of $\tilde{\mathbf{B}}^d$ corresponding to sub-carrier $i$. By again computing the QR decomposition of $(\mathbf{Q}^d)^H$ and after doing some mathematical manipulations, one can get (see (8) of [37] for the details)

$$\hat{d}_{ik}^d = \mathbf{h}_{ik}^H \tilde{\mathbf{A}} \tilde{\mathbf{B}} \tilde{\mathbf{B}}_i^d \mathbf{d}_i + n_{ik}, \quad \forall i,k \tag{8}$$

where $\tilde{\mathbf{A}} = [\mathbf{G}, \ \tilde{\mathbf{G}}]^H$, $\mathbf{G} = \text{diag}(g_1, g_2, \cdots, g_{r_t})$, $\tilde{\mathbf{G}} \in \mathcal{C}^{r_t \times (N-r_t)}$ and $\tilde{\mathbf{B}} \in \mathcal{C}^{r_t \times r_t}$, with each elements of $\mathbf{G}$ and $\tilde{\mathbf{G}}$ has a maximum amplitude of 2. Note that for a massive MIMO application, low complexity QR decomposition algorithms can be applied [38].

## IV. HYBRID BEAMFORMING DESIGNS

In this section, we describe the design of analog and digital precoder matrices of (3). As in the conventional DB, the entries of $\mathbf{A}$ and $\mathbf{B}_i$ of this equation can be optimized by considering different design objectives such as sum rate maximization, SINR balancing etc. However, the rows of $\mathbf{B}_i$ depends on the available number of RF chains $N_a$ which is fixed a priori in the production stage of the BS. Furthermore, since $\mathbf{A}$ is realized using electronic components (i.e., PSs), the number of PSs $N_{PS}$ to realize $\mathbf{A}$ is again fixed during the production stage of the BS.

This section determines $N_a$ and $N_{PS}$ such that the HB design is able to maintain the same performance as that of the DB (which uses $N$ RF chains) for any design criteria. Indeed, this can be met whenever $N_a$ and $N_{PS}$ are determined while ensuring $\hat{d}_{ik} = \hat{d}_{ik}^d$ (i.e., the received signal with HB is the same as that of the DB) without considering a design criteria. In the following, we examine **P1** by exploiting this idea. Specifically, first we provide the proposed HB design where the digital precoding part of Fig. 1 is realized using microprocessors whereas, its analog precoding employs infinite resolution PSs (**P1**$_A$). Then, we extend this result to handle **P1**$_B$.

### A. Hybrid Beamforming Design for **P1**$_A$

This section discusses the proposed HB design for **P1**$_A$. One can notice that (3) and (8) have the same mathematical structure. Hence, one may think of directly setting $\mathbf{A} = \tilde{\mathbf{A}}$ and $\mathbf{B}_i = \tilde{\mathbf{B}} \tilde{\mathbf{B}}_i^d$ to address **P1**$_A$. However, since the amplitude of each of the elements of $\tilde{\mathbf{A}}$ is not necessarily one, it is not clear how one can realize this matrix using PSs only. Hence, such direct plug-in will not help to design the HB architecture. For this reason, the authors of [6] first come up with a novel and clever method to represent "any" vector $\mathbf{x} \in \mathcal{C}^{N \times 1}$ as $\mathbf{x} = \mathbf{Wz}$ (see Theorem 1 of [6]), where $\mathbf{W} \in \mathcal{C}^{N \times 2}$ and $\mathbf{z} \in \mathcal{C}^{2 \times 1}$ with $|\mathbf{W}_{(i,j)}|^2 = 1, \forall i, j$ which leads them to conclude that the performance of any DB can be achieved with the HB if the number of RF chains are at least two times that of the number of data streams (i.e., two times the rank of $\mathbf{B}^d$) [5], [6]. Now if we utilize this technique to our HB architecture, $2r_t$ RF chains and $2r_t N$ digitally controlled PSs (DCPSs) are needed to achieve the same performance as that of the DB design.

In the following, we propose new and simple method to reduce the number of RF chains and DCPSs compared to [6]. To this end, we consider the following theorem.

*Theorem 1*: Given any real number $x$ with $-2 \leq x \leq 2$, it can be shown that

$$x = e^{j \cos^{-1}(\frac{x}{2})} + e^{-j \cos^{-1}(\frac{x}{2})} \tag{9}$$

$$jx = e^{j \sin^{-1}(\frac{x}{2})} + e^{j(\pi - \sin^{-1}(\frac{x}{2}))} \tag{10}$$

where $j = \sqrt{-1}$.

*Proof:* When $-1 \leq \frac{x}{2} \leq 1$, we will have

$$e^{j \cos^{-1}(\frac{x}{2})} + e^{-j \cos^{-1}(\frac{x}{2})} =$$

$$\cos(\cos^{-1}(\frac{x}{2})) + j \sin(\cos^{-1}(\frac{x}{2})) + \cos(-\cos^{-1}(\frac{x}{2})) + j \sin(-\cos^{-1}(\frac{x}{2})) = x.$$

Similar to this expression, one can prove that $e^{j \sin^{-1}(\frac{x}{2})} + e^{j(\pi - \sin^{-1}(\frac{x}{2}))} = jx$. ∎

The $(m,n)$th element of $\tilde{\mathbf{A}}$ can also be rewritten as $\tilde{a}_{mn} e^{j \phi_{mn}}$, where $0 \leq \tilde{a}_{mn} \leq 2$. By applying (9) of *Theorem 1*, we can express $\tilde{\mathbf{A}}_{(m,n)}$ as

$$\tilde{\mathbf{A}}_{(m,n)} = \tilde{a}_{mn} e^{j \phi_{mn}} = e^{j(\cos^{-1}(\frac{\tilde{a}_{mn}}{2}) + \phi_{mn})} + e^{-j(\cos^{-1}(\frac{\tilde{a}_{mn}}{2}) - \phi_{mn})}. \tag{11}$$

From this equation, we can notice that each element of $\tilde{\mathbf{A}}$ can be equivalently expressed as a sum of two DCPSs. As the maximum number of non-zero elements of $\tilde{\mathbf{A}}$ is $r_t(N - r_t + 1)$ (i.e., from (8)), the solution obtained in DB can be achieved by employing $2r_t(N - r_t + 1)$ DCPSs and $r_t$ RF chains. This leads us to get the HB architecture of Fig. 2.(a) by setting

$$\mathbf{A} = \tilde{\mathbf{A}}, \quad \mathbf{B}_i = \tilde{\mathbf{B}} \tilde{\mathbf{B}}_i^d. \tag{12}$$

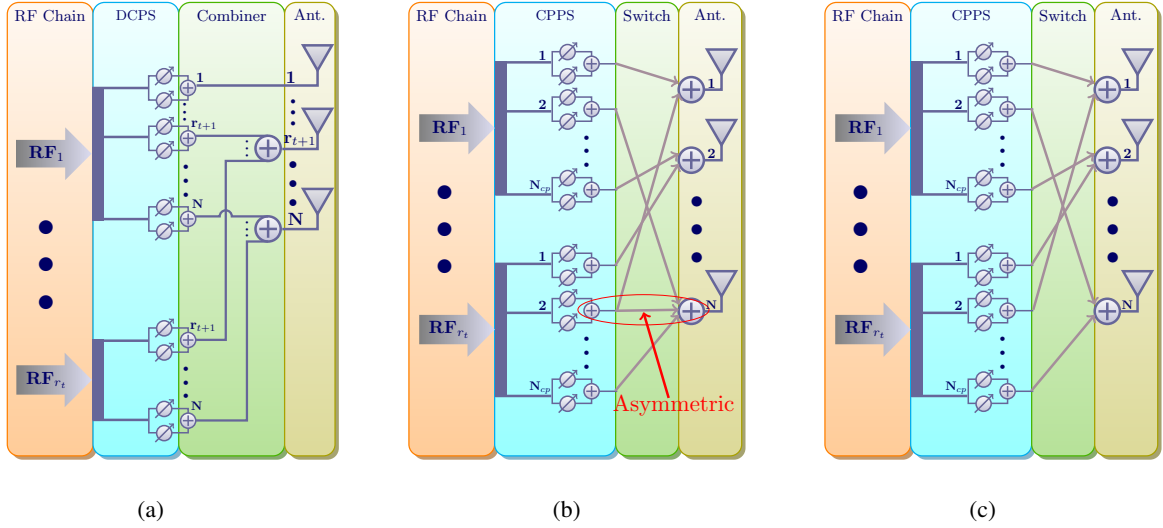

Fig. 2: Detailed HB design with: (a) DCPSs (b): CPPSs and Switches (Asymmetric) (c): CPPSs and Switches (Symmetric).

From this result, the following ideas can be noted: The mathematical manipulation on $\hat{d}^d_{ik}$ "alone" does not bring new result. However, by utilizing $\hat{d}^d_{ik}$ and *Theorem 1*, we are able reduce the number of RF chains (by half) and DCPSs (slight difference) compared to [6]. On the other hand, the result of this theorem also helps us to come up with a practical scheme to realize **A** with limited number of CPPSs which is discussed in the next Section[2].

We would like to emphasize here that the DCPSs of this section are assumed to have infinite resolution which may not be realistic. Thus, the result of this section can be used as a benchmark for "future theoretical results" (or other practical designs) in the HB research.

### B. Realizing **A** with Limited Number of CPPSs ($\mathbf{P1}_B$)

The HB design approach of the above subsection has limitations which arises on how to realize **A** using practical PSs. This subsection provides a simple approach to realize **A** just by using CPPSs and switches only (i.e., $\mathbf{P1}_2$). In this regard, it is considered that $N_{cp}$ pairs of CPPSs are shared by each RF chain and all BS antennas with the help of switches as shown in Fig. 2.(b).

---

[2]We have learned that recently [39] ensures the same performance as that of the DB for single user case. The authors of [39] basically come up with the same result as (9) which is proven in a different method. Thus, the work of [39] can be used for the setup of the current paper. However, still it is not clear how to apply the approach of [39] to examine $\mathbf{P1}_B$ and $\mathbf{P2}$.

In fact, any switch can be represented by 0 (disconnected) or 1 (connected). By denoting the switching matrix between the CPPSs of the $i$th RF chain and antennas as $\mathbf{S}^i$, the analog precoding corresponding to the $i$th RF chain of the HB can be expressed as $\mathbf{S}^i \mathbf{d}^i_{cp}$, where $\mathbf{S}^i \in \{0,1\}$ is an $N \times N_{cp}$ sized switching matrix and $\mathbf{d}^i_{cp}$ is an $N_{cp}$ sized vector whose entries are the scalar values corresponding to a pair of CPPSs. In this design, it is assumed that each pair of CPPSs can be connected to a maximum of $\tilde{L}$ antennas. Furthermore, each antenna will receive signals from a maximum of $\bar{L}$ CPPSs corresponding to each RF chain. These constraints are equivalent to ensuring that the maximum sum of each column (row) of $\mathbf{S}^i$ to be $\tilde{L}(\bar{L})$ as shown in Fig. 2.(b). With these constraints, one approach of examining $\mathbf{P1}_B$ is by first solving

$$\min_{\mathbf{S}^i, \mathbf{d}^i_{cp}, \tilde{L}, \bar{L}} \sum_{i=1}^{r_t} |\mathbf{S}^i \mathbf{d}^i_{cp} - \tilde{\mathbf{A}}_{(:,i)}|^2 \ \text{s.t} \ \sum \mathbf{S}^i_{(:,m)} \leq \tilde{L}, \ \sum \mathbf{S}^i_{(m,:)} \leq \bar{L}, \ \mathbf{S}^i_{(m,n)} \in \{0,1\} \quad (13)$$

and then setting $\mathbf{A}_{(:,i)} = \mathbf{S}^i \mathbf{d}^i_{cp}$ where $\tilde{\mathbf{A}}_{(:,i)}$ is as defined in (12).

The above explanation suggests that each entry of $\mathbf{d}^{cp}_i$ is a scalar value corresponding to a pair of CPPSs. Furthermore, $\tilde{L}$ and $\bar{L}$ correspond to the number of available switches. Since $\mathbf{d}^i_{cp}$, $\tilde{L}$ and $\bar{L}$ are determined a priori in the production stage of PSs and switches, they cannot be optimized for each realizations of $\tilde{\mathbf{A}}_{(:,i)}$. The above problem can therefore be solved by applying a two step approach; determination of $\mathbf{d}^i_{cp}$, $\tilde{L}$ and $\bar{L}$ (for general $\tilde{\mathbf{A}}_{(:,i)}$), and optimization of $\mathbf{S}^i$ (for each realization of $\tilde{\mathbf{A}}_{(:,i)}$) while keeping $\mathbf{d}^i_{cp}$, $\tilde{L}$ and $\bar{L}$ constant. To this end, we consider two cases where the first (second) case allows the switching matrices $\mathbf{S}^i$ to be designed by enabling asymmetric (symmetric) signal flow in between RF chains and antennas.

*1) Case 1: Asymmetric signal flow:* For such a case, we suggest to examine the above problem such that $|\mathbf{A}_{(:,i)} - \tilde{\mathbf{A}}_{(:,i)}|^2 \leq 2\epsilon^2 N$ is ensured for arbitrary desired error tolerance $\epsilon > 0$. One approach of maintaining this inequality is by designing $\mathbf{d}^i_{cp}$, $\mathbf{S}^i$, $\tilde{L}$ and $\bar{L}$ while ensuring $|\Re\{\mathbf{A}_{(j,i)}\} - \Re\{\tilde{\mathbf{A}}_{(j,i)}\}| \leq \epsilon$ and $|\Im\{\mathbf{A}_{(j,i)}\} - \Im\{\tilde{\mathbf{A}}_{(j,i)}\}| \leq \epsilon$. To this end, let us consider a simple example to illustrate our solution approach when $\epsilon = 10^{-2}$.

The accuracy $\epsilon = 10^{-2}$ means that a number in between 0 and 1 is represented by 2 decimal places only. For example, $0.1416$ is represented as $0.14$. Furthermore, with this accuracy level, any number in between 0 and 1 can be represented as a sum of two values taken from $\mathcal{F} = [0.1, 0.2, \cdots, 1]$ and $\tilde{\mathcal{F}} = [0.00, 0.01, 0.02, \cdots, 0.09]$ (for instance, $0.14 = 0.1 + 0.04$).

This shows that for an accuracy of $\epsilon = 10^{-p}$, only $10p$ numbers are required to represent any scalar value in between 0 and 1. We employ this number representation in our HB design. That is, for the accuracy level of $\epsilon = 10^{-2}$, using the result of *Theorem 1*, the following CPPSs are required to ensure $|\Re\{\mathbf{A}_{(j,i)}\} - \Re\{\tilde{\mathbf{A}}_{(j,i)}\}| \leq \epsilon$ in each RF chain

$$\text{Real} = \begin{cases} \pm \cos^{-1}(-1.00), \cdots, \pm \cos^{-1}(-0.10), \pm \cos^{-1}(0.10), \cdots, \pm \cos^{-1}(1.00) \\ \pm \cos^{-1}(-0.09), \cdots, \pm \cos^{-1}(-0.01), \pm \cos^{-1}(0.01), \cdots, \pm \cos^{-1}(0.09) \end{cases},$$

$$\text{Imag} = \begin{cases} \sin^{-1}(\pm 0.10), \cdots, \sin^{-1}(\pm 1.00), \pi - \sin^{-1}(\pm 0.10), \cdots, \pi - \sin^{-1}(\pm 1.00) \\ \sin^{-1}(\pm 0.01), \cdots, \sin^{-1}(\pm 0.09), \pi - \sin^{-1}(\pm 0.00), \cdots, \pi - \sin^{-1}(\pm 0.09). \end{cases} \quad (14)$$

As discussed above, each of the real (complex) entries of $\tilde{\mathbf{A}}_{(:,i)}$ are in the range of $[-2, 2]$. Thus, to realize each of these entries with accuracy $\epsilon = 10^{-2}$, four CPPSs taken from the above sets are required. For example, if the real part of the $n$th element of $\tilde{\mathbf{A}}_{(:,i)}$ is 1.64, it can be represented by using four PSs (i.e., $e^{j\cos^{-1}(0.80)} + e^{-j\cos^{-1}(0.80)} + e^{j\cos^{-1}(0.02)} + e^{-j\cos^{-1}(0.02)}$. Using this result, it can be shown that to achieve $10^{-p}$ accuracy, "approximately" $\lceil 40p \rceil$ shared pairs of CPPSs are required per each RF chain. As $\mathbf{A}_{(j,i)}$ can be realized by at most $4p$ pairs of PSs where each of these pairs are uniquely obtained from at most 10 CPPSs, the complexity of searching these $4p$ pairs of CPPSs is negligible. Also, the phases of these CPPSs are not necessarily spaced uniformly (for example, $\cos^{-1}(0.10), \cos^{-1}(0.20)$ and $\cos^{-1}(0.30)$ are $84.2°, 78.4°, 72.5°$ and $66.4°$, respectively). Thus, with this design, we have $N_{cp} = 40p$, $\tilde{L} = N$, $\bar{L} = 4p$, $\mathbf{d}_{cp}^i = \mathbf{d}_{cp}^k = \mathbf{d}_{cp}, \forall i, k$ are taken from pairs of CPPSs as in (14) and computing $\mathbf{S}^i$ is straightforward.

In this design, each pair of CPPSs can be connected to a maximum of $N$ antennas. In such a case, there could be a scenario where a pair of CPPSs can be connected to one antenna only whereas, the other pairs are connected to all the $N$ antennas (i.e., $\sum \mathbf{S}^i_{(m,:)} \leq \tilde{L} = N$). Such a phenomena may lead to asymmetric signal flow in the architecture as shown in Fig. 2.(b) which may not be desirable in practice[3]. This motivates us to consider Case 2 in the following.

*2) Case 2: Symmetric signal flow:* For this case, the determination of $\mathbf{d}_{cp}^i$, $\tilde{L}$ and $\bar{L}$, and optimization of $\mathbf{S}^i$ while ensuring a prescribed error tolerance $\epsilon$ is not trivial. In the following, we provide simple method to address the problem considered in this case. To this end, we utilize $\mathbf{d}_{cp}$ and $\bar{L}$ as in Case 1 (i.e., $N_{cp} = 40p$, $\bar{L} = 4p$) but modify $\tilde{L}$ as $\frac{N}{\bar{L}}$. With these settings, we round-off each of the significant digits of the real (imaginary) components of $\mathbf{A}_{(:,i)}$ such that the rounded-off vector will lie to its corresponding significant digits[4]. By doing so, we have observed negligible performance loss compared to Case 1 (see Section VII-D for the details).

---

[3]In scenario where asymmetric signal flow does not have an impact, the approach discussed up to now can be utilized.

[4]Here $p$ is selected such that $\tilde{L} \geq 2$ is satisfied (i.e., at least one switch for each of the real and imaginary term of $\mathbf{A}_{(j,i)}$).

We would like to recall here that the design approach of this subsection is "only a particular HB architecture" and hence it may not be the global optimal design. By utilizing the proposed design, however, we achieve the following advantages compared to those of the existing ones:

1) The existing HB approaches (for example [17], [18], [23], [29]) utilize quantized DCPSs whereas, the proposed approach employs CPPSs. This puts our design to be advantageous over those of the existing ones as the price and energy consumption of DCPSs are much higher than those of CPPSs especially when the DCPSs have high resolution [40].

2) We are able to provide an insight on the relation between the number of CPPSs and the accuracy of analog beamforming matrix which is valid for any design criteria and channel when asymmetric signal flow is allowed. This helps us to design the analog beamforming matrix while maintaining symmetric signal flow which is practically useful.

Wireless communications channels have a non zero coherence time $T_c$ where the channel is assumed to be almost constant. Thus, both the digital and analog beamforming matrices may need to be updated every $T_c$ seconds (for example, $T_c \approx 0.5$ millisecond in long term evolution (LTE) network [4], [41]). Hence, for microwave frequency bands, the proposed HB can utilize electronic switches which needs to be updated once every $T_c$ [42]. Furthermore, mmWave switches are capable of switching at a fraction of nanosecond speeds where smart switches have also been used for 60 GHz application at the receiver side in [43]. On the other hand, according to the detailed study of [40], the price and energy consumptions of switches are insignificant compared to those of PSs [40], [44]. In some cases, reasonable performance can be obtained just by designing the beamforming matrices based on the long term channel statistics as in [45] where the beamforming matrices are kept constant for the duration much larger than $T_c$. For these reasons, we believe that the introduction of switches will incur negligible delay, cost and energy consumption in the proposed HB design [40], [44].

Note that we have provided three different hybrid architecture implementation aspects where each of them requires different sets of PSs and/or switches. As will be demonstrated in the simulation section, the performances of all these architectures are very close to each other. As mentioned above, the price of digitally controlled PSs is much higher than that of PSs having fixed phases. On the other hand, the digitally controlled switches can be realized like in [43] and this design can be customized to have either fully connected (partially) connected HB architecture as suggested in [33]. Due to these reasons, we believe that the architecture proposed in Fig. 2.(c) is cost effective, simple to implement and can achieve the desired performance. Having said this, however, the detailed comparison of this design compared to those of Fig. 2.(a), Fig. 2.(b), and the existing hybrid architectures in terms of cost, energy consumption and performance requires significant effort and time, and it is left for future research.

The HB design approach discussed in this section employs $r_t = \text{rank}(\mathbf{B}^d)$ RF chains. However, for an arbitrary channel matrix of all sub-carriers and $K$, the number of deployed RF chains ($N_a$ in Fig. 1) may be less than that of the rank of $\mathbf{B}^d$. In such a case, the DB cannot be implemented using the proposed HB architecture. The following section provides the proposed user scheduling and sub-carrier allocation algorithm while ensuring $\text{rank}(\mathbf{B}^d) \leq N_a$.

## V. Proposed User Scheduling and Sub-carrier Allocation (P2)

This section provides the proposed user scheduling. One can understand from the above section that the solution of the scheduler can be realized using the proposed HB design (i.e., with the desired accuracy) if the HB architecture has $N_a$ RF chains and $\text{rank}(\mathbf{B}^d) \leq N_a$. Hence, one can examine the scheduling problem to optimize $\mathbf{B}^d$ while introducing this constraint. In practice a scheduler is usually designed to optimize some performance criteria. To this end, we examine maximization of the sum rate of all sub-carriers with a per sub-carrier power constraint as

$$\max_{\mathbf{B}_i^d} \sum_{i=1}^{N_f} \sum_{k=1}^{K_i} \log(1+\gamma_{ik}), \quad \text{s.t } \text{tr}\{(\mathbf{B}_i^d)^H \mathbf{B}_i^d\} \leq P_i, \quad \text{rank}(\mathbf{B}^d) \leq N_a \qquad (15)$$

where $K_i(P_i)$ is the number of UEs served (available power) in sub-carrier $i$ and $\gamma_{ik}$ is SINR of the $k$th UE in sub-carrier $i$ (i.e., $\gamma_{ik} = \frac{|\mathbf{h}_{ik}^H \mathbf{b}_{ik}^d|^2}{\sum_{j \neq k} |\mathbf{h}_{ik}^H \mathbf{b}_{ij}^d|^2 + \sigma^2}$ with $\mathbf{B}_i^d = [\mathbf{b}_{i1}^d, \mathbf{b}_{i2}^d, \cdots, \mathbf{b}_{iK}^d]$).

In [46], it is shown that the ZF precoding approach together with user scheduling achieves the capacity region of a multiuser system when the total number of scheduled users $K_t$ are very large. Furthermore, in a massive MIMO setup with sufficient number of scatterers, a simple precoding approach such as ZF precoding technique can achieve the optimal sum rate [4]. Due to these reasons, we utilize ZF precoding to design $\mathbf{B}_i^d$ of problem (15).

When $N_a = N$ (i.e., DB scenario), the rank constraint of (15) is satisfied implicitly and the above problem can be examined independently for each sub-carrier as

$$\max_{\mathbf{B}_i^d} \sum_{k=1}^{K_i} \log(1+\gamma_{ik}) \triangleq f(\mathbf{B}_i^d), \quad \text{s.t } \text{tr}\{(\mathbf{B}_i^d)^H \mathbf{B}_i^d\} \leq P_i, \quad \forall i. \qquad (16)$$

However, when $N_a < N$, the solution of (16) may not necessarily satisfy the rank constraint of (15). In the following, we discuss the proposed user scheduling and sub-carrier allocation algorithm to solve (15) which is summarized in **Algorithm I**.

**Algorithm I**: User scheduling and sub-carrier allocation algorithm.

**Input**: Users to schedule $\{1, 2, \cdots, K_t\}$, $N_f$, $K_i$ and $N_a$.

**Phase I**:

  for $i = 1 : N_f$ do

    for $n = 1 : K_i$ do

1) Set $\mathbb{K}_{im} = \mathbb{K}_i \cup \{m\}$, $\forall m \in \mathbb{K}_{ti}$, where $\cup$ denotes union.
2) Compute $f_{im}(\mathbf{B}_{im}^d)$, where $f_{im}(\mathbf{B}_{im}^d)$ is the objective function of (16) with $\mathbb{K}_{im}$ users.
3) Compute $\tilde{m}_i = \arg\max\{f_{im}(\mathbf{B}_{im}^d), \forall m\}$
4) Set $\mathbf{B}_i^d = \mathbf{B}_{i\tilde{m}_i}^d$ and $f(\mathbf{B}_i^d)^{new} = f_{i\tilde{m}_i}(\mathbf{B}_{i\tilde{m}_i}^d)$
5) **if** $f(\mathbf{B}_i^d)^{new} \geq f(\mathbf{B}_i^d)^{old}$ **then**
   - Update $\mathbb{K}_i = \mathbb{K}_i \cup \{\tilde{m}_i\}$, $\mathbb{K}_{ti} = \mathbb{K}_{ti} \backslash \{\tilde{m}_i\}$ and $f(\mathbf{B}_i^d)^{old} = f(\mathbf{B}_i^d)^{new}$.
6) **else**
   - Break
7) **end if**

    end for

  end for

8) Stack the precoders $\mathbf{B}^d = [\mathbf{B}_1^d, \mathbf{B}_2^d, \cdots, \mathbf{B}_{N_f}^d]$ and

**if** $\text{rank}(\mathbf{B}^d) \leq N_a$ **then**
- Employ this $\mathbf{B}^d$ as the HB precoder.

**else**
- Go to **Phase II**

**end if**

**Phase II**:

1) Sort $f(\mathbf{B}_i^d), \forall i$ in decreasing order $f(\mathbf{B}_1^d) \geq f(\mathbf{B}_2^d) \geq, \cdots, \geq f(\mathbf{B}_{N_f}^d)$.
2) Compute $\mathbf{T} = [\mathbf{B}_1^d, \cdots, \mathbf{B}_{\tilde{S}}^d]$, where $\tilde{S}$ is minimum number of sub-carriers with $\text{rank}(\mathbf{T}) \geq N_a$.
3) Compute $\text{SVD}(\mathbf{T}) = \mathbf{U}\mathbf{\Lambda}\mathbf{V}^H$ with decreasing order of the diagonal elements of $\mathbf{\Lambda}$.
4) Set $\mathbf{Q}^d$ of (7) as the first $N_a$ columns of $\mathbf{U}$.
5) For fixed $\mathbf{Q}^d$, perform **Phase I** for the system (17) and set $\mathbf{B}_i^d$ as $\mathbf{B}_i^d = \mathbf{Q}^d \bar{\mathbf{B}}_i^d$.

**Output**: The precoders of all sub-carriers $\mathbf{B}_1^d, \mathbf{B}_2^d, \cdots, \mathbf{B}_{N_f}^d$ and corresponding scheduled users.

As we can see, our algorithm employs two phases which are explained as follows. In the first phase, we examine (15) by dropping its rank constraint. This rank relaxed problem (i.e., (16)) is solved iteratively by increasing its sum rate and number of served users simultaneously for each sub-carrier. Then, we compute $\text{rank}(\mathbf{B}^d)$ and, if $\text{rank}(\mathbf{B}^d) \leq N_a$, as the constraint of (15) is satisfied, we consider this $\mathbf{B}^d$ as our hybrid precoder. However, if $\text{rank}(\mathbf{B}^d) > N_a$, the constraint of (15) is violated and we will execute the second phase. In this phase, first, we compute $\mathbf{Q}^d$ from the singular value decomposition (SVD) of the precoders of the first $\tilde{S}$ sub-carriers having the maximum sum rate, where $\tilde{S}$ is the minimum number of sub-carriers ensuring $\text{rank}([\mathbf{B}_1^d, \mathbf{B}_2^d, \cdots, \mathbf{B}_{\tilde{S}}^d]) \geq N_a$. Then, for fixed $\mathbf{Q}^d$, we re-express (7) as

$$\hat{d}_{ik}^d = \tilde{\mathbf{h}}_{ik}^H \bar{\mathbf{B}}_i^d \mathbf{d}_i + n_{ik}, \quad \forall i, k \tag{17}$$

where $\tilde{\mathbf{h}}_{ik}^H = \mathbf{h}_{ik}^H \mathbf{Q}^d$. Finally, we perform **Phase I** for the system (17) and set $\mathbf{B}_i^d$ as $\mathbf{B}_i^d = \mathbf{Q}^d \bar{\mathbf{B}}_i^d$.

From this explanation, we can understand that a given user may or may not be scheduled to use all of the available sub-carriers. We would like to mention here that **Algorithm I** can also be extended straightforwardly for other design criteria and precoding method.

**Convergence of Algorithm I**: As can be seen from **Algorithm I**, the proposed scheduling algorithm employs a two step approach where the number of UEs is increased sequentially. From step 5 of this algorithm, one can notice that the UEs are selected sequentially while ensuring a non decreasing total sum rate. Furthermore, as the BS has finite available power, the sum rate achieved by the proposed algorithm is finite. For these reasons, the proposed algorithm is guaranteed to converge to a finite total sum rate. However, **Algorithm I** may not necessarily converge to the global optimal solution, and we believe that the development of global optimal user scheduling and sub-carrier allocation algorithm under the proposed hybrid beamforming design is not trivial, and it is still an open research problem.

## VI. PERFORMANCE ANALYSIS

In this section we provide performance analysis of the proposed user scheduling and sub-carrier allocation algorithm. Massive MIMO system can be realized both at the microwave and mmWave frequency bands. Recently several field measurements are conducted to examine the characteristics of massive MIMO channels. For the practically relevant number of antenna elements, it has been demonstrated that the channel covariance matrices of each UE experiences higher rank in a typical outdoor environment in a microwave massive MIMO system. These results also suggest that despite the statistical difference between the measured channels and the i.i.d. channels, most of the theoretical conclusions made under the independence assumption (i.e. i.i.d Rayleigh fading channel) are still valid for the real massive MIMO channels in the microwave frequency bands [47], [48]. On the other hand, the utilization of ULA antenna array patters are commonly employed and justified for wireless applications where mmWave (microwave) frequency bands are fairly represented by very low (high) scatterers [17], [45], [49], [50]. These reasons motivate us examine the performance of the proposed user scheduling and sub-carrier allocation algorithm for the Rayleigh fading and ULA channel models. By combining the ZF precoding and **Algorithm I**, problem (15) can be solved and realized by the following three possible approaches.

1) **Antenna Selection Beamforming Approach:**

   When the beamforming matrix $\mathbf{B}_i^d$ has effective size $N_a \times K_i$ matrix (i.e., when the remaining entries of $\mathbf{B}_i^d$ are set to 0 a priory), the rank constraint of (15) is satisfied implicitly. Thus, for such a setting, this problem can be solved independently for each sub-carrier just by employing the ZF precoding and **Phase I** of **Algorithm I**. As this approach implicitly selects $N_a$ antennas from $N$ available antennas, we refer to this approach as an antenna selection beamforming (ASB). We would like to mention here that such an approach is widely known in the existing literature [51]. Hence, the ASB approach can be treated as an existing approach.

2) **Proposed Hybrid Beamforming Approach:**

   In this approach, we utilize the proposed HB architecture of Fig. 1. Here we apply the ZF precoding to design the precoders $\mathbf{B}_i^d$ and **Algorithm I** to schedule the served users and sub-carriers. We refer to this as the proposed HB approach.

3) **Digital Beamforming Approach:**

   The upper bound solution of problem (15) is achieved when we have $N$ number of RF chains which corresponds to the conventional DB approach.

In the following, we analyze the performances of these three approaches for the Rayleigh fading and ULA channel models.

### A. Rayleigh Fading Channel

In this subsection, we examine the above approaches by assuming that the channel coefficients $\tilde{\mathbf{h}}_k^H, \forall k$ of (4) are i.i.d Rayleigh fading.

*Lemma 1*: Under ZF beamforming, Rayleigh fading channel $\tilde{\mathbf{h}}_k^H$ and large $K_t$, we can have

$$R_i^{HB} \geq R_i^{ASB} \text{ when } \mathbb{K}_i^{HB} = \mathbb{K}_i^{ASB}$$

where $\mathbb{K}_i^{ASB}(R_i^{ASB})$ and $\mathbb{K}_i^{HB}(R_i^{HB})$ are the served set of users (achieved sum rate) in sub-carrier $i$ using the existing ASB and proposed HB approaches, respectively.

   *Proof:* See Appendix A. ∎

From *Lemma 1*, we understand that the proposed HB achieves the same sum rate as that of the DB one when $\mathbb{K}_i^{HB} = \mathbb{K}_i^{DB}$. However, in general, the set of served users (obtained by **Algorithm I**) of the HB and DB approaches may not be necessarily the same for all channel realizations.

This motivates us to examine the performances of the aforementioned three approaches for the case where $\mathbb{K}_i^{ASB} \neq \mathbb{K}_i^{HB} \neq \mathbb{K}_i^{DB}$ for some $i$. For such a case, we are not able to quantify the relation between $R_i^{ASB}$, $R_i^{HB}$ and $R_i^{DB}$ for each channel realization. Thus, here we compare the performances of these three approaches by examining their achieved average rates under ZF beamforming with equal power allocation strategy as follows.

*Theorem 2*: Under ZF beamforming with equal power allocation, $P_i = P, K_i = K$ and a unit variance i.i.d Rayleigh fading channel $\tilde{\mathbf{h}}_k^H$, we can have the following average rates.

$$\mathrm{E}\{R^{ASB}\} \leq KN_f \log_2\left(1 + \frac{P}{K}\mathrm{E}\{\chi_{max}^{N_a-K+1}(K_g)\}\right)$$

$$\mathrm{E}\{R^{DB}\} \leq KN_f \log_2\left(1 + \frac{P}{K}\mathrm{E}\{\chi_{max}^{N-K+1}(K_g)\}\right) \quad (18)$$

$$\mathrm{E}\{R^{HB}\} \leq K\tilde{S} \log_2\left(1 + \frac{P}{K}\mathrm{E}\{\chi_{max}^{N-K+1}(K_s)\}\right) + K(N_f - \tilde{S}) \log_2\left(1 + \frac{P}{K}\mathrm{E}\{\chi_{max}^{N_a-K+1}(K_g)\}\right)$$

where $\tilde{S} \geq 1$, $K_g = \lceil \frac{K_t}{K} \rceil$, $K_s = \lceil \frac{K_t N_f}{K N_a} \rceil$ and the notation $\mathrm{E}\{\chi_{max}^M(L)\}$ denotes the expected value of the maximum of $L$ independent Chi-square distributed random variables each with $M$ degrees of freedom[5].

*Proof:* See Appendix B. ∎

## B. Uniform Linear Array (ULA) Channel

From the proof of *Theorem 2*, we can observe that the proposed HB approach achieves lower average sum rates than that of the DB approach. This performance loss occurs due to the rank constraint of $\mathbf{B}^d$ of (15). For the ZF precoding of this paper, $\mathbf{B}^d$ has the same rank as that of the combined channels of all users. Thus, the proposed HB approach achieves the same performance as that of the DB one if the combined channel of all of the $K_t$ users has approximately a maximum rank of $N_a$. In this regard, we consider the following lemma.

*Lemma 2*: When $\tilde{d} = \frac{\lambda}{2}$ and the AOD of the $K_t$ users satisfy $\sin(\theta_{km}) \in n\sin(\theta)[-\frac{1}{2N}, \frac{1}{2N}]$, $n = 1, 2, \cdots, N_a$, where $\theta$ is an arbitrary angle, we can achieve

$$\mathbb{K}_i^{HB} = \mathbb{K}_i^{DB} \text{ and } R_i^{HB} = R_i^{DB}, \ \forall i.$$

*Proof:* See Appendix C. ∎

---

[5]For the simulation, we employ simple trapezoid numerical integration approach of Matlab to compute $\mathrm{E}\{\chi_{max}^M(L)\}$. As will be demonstrated in the simulation section, the bound derived in this theorem is tight.

## VII. SIMULATION RESULTS

This section presents simulation results. We have used $N_f = 64$, $L_p = 8$ (i.e., 8 tap channel), $\rho_k = 1, \forall k$ and $K_{max} = 8$. The signal to noise ratio (SNR) which is defined as $SNR = \frac{N_f P}{K_{max} \sigma^2}$ is controlled by varying $P_i = P$ while keeping the noise power at 1mW. We have used the Rayleigh and ULA fading channel models as defined in Section III-A. All of the plots are generated by averaging over 1000 channel realizations and ASR denotes average sum rate. In Sections VII-A - VII-C, the analog beamforming part of the proposed HB design approach is designed by considering asymmetric signal flow with $\epsilon = 10^{-1}$.

### A. Rayleigh Fading Channel

In this subsection, we provide simulation results for the scenario where the channel (4) is taken from i.i.d Rayleigh fading channel model.

*1) Verification of Theoretical Rates:* In this simulation, we examine the tightness of the upper bound rates given in (18) under equal power allocation policy. To this end, we take $N = 64$, $N_a = 16$, $K_i = K_{max}$ and $K_t = 8$. Fig. 3 shows the rates achieved by simulation and theory. As can be seen from this figure, the bound derived in (18) is very tight. Furthermore, as expected the rate achieved by the proposed HB approach is higher than that of the existing ASB approach, and superior performance is achieved by the DB approach.

*2) Effect of Power Allocation and Number of Users ($K_i$):* As can be observed from Section VI, the theoretical average sum rate expressions of (18) is derived by assuming that $K_i$ is fixed a priori and Fig. 3 is plotted for fixed $K_i = K_{max}$. However, when we employ **Algorithm I**, the number of served users per sub-carrier is updated adaptively. Hence the number of served users per sub-carrier may vary from one channel realization to another. Furthermore, from fundamentals of MIMO communications, ZF precoding with water filling power allocation achieves better performance than that of the equal power allocation. This simulation demonstrates the joint benefits of the ZF precoding with water filling power allocation and **Algorithm I** (i.e., choosing $K_i$ adaptively). To this end, we set $K_i \leq K_{max}$, $K_t = 16$ and $N = 64$. Fig. 4 shows the performances of the existing ASB, proposed HB and DB approaches for these parameter settings. As we can see from this figure, for all approaches, performing power allocation with adaptive $K_i$ is advantageous which is expected[6]. In the subsequent simulations, we employ ZF precoding with water filling power allocation and **Algorithm I** (i.e., the number of served users of sub-carrier $i$ $K_i \leq K_{max}$ is chosen adaptively).

---

[6]Note that the complexity of water filling power allocation is almost the same as that of the equal power allocation.

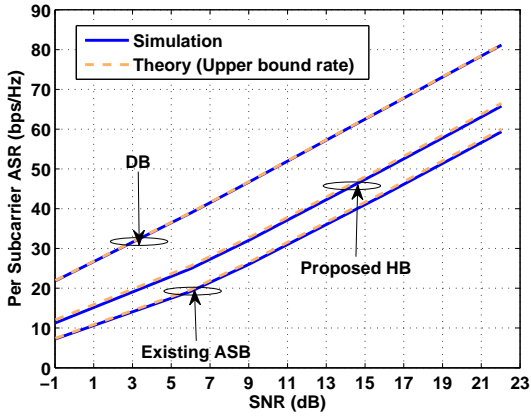
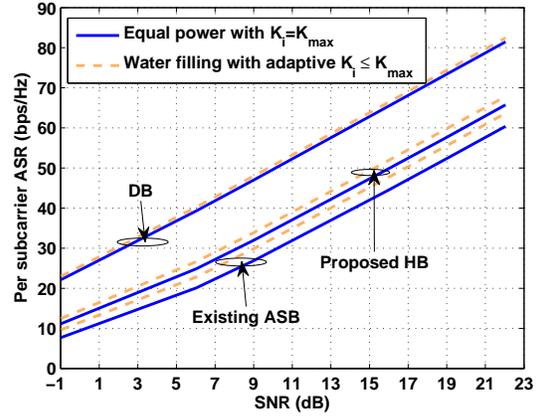

Fig. 3: Comparison of theoretical and simulated ASR of the existing ASB, proposed HB and DB approaches under ZF precoding and equal power allocation.

Fig. 4: Comparison of ASRs achieved by ZF precoding with equal power and $K_i = K_{max}$ versus ZF precoding with water filling power allocation and adaptive $K_i$.

*3) Comparison of Proposed HB and Existing ASB Approaches:* In this simulation, we examine and compare the performances of the proposed HB and existing ASB approaches for different parameter settings. Fig. 5 shows the average sum rate achieved by these approaches for different SNR and $K_t$. From this figure, we can observe that increasing $K_t$ increases the average sum rate of both approaches (for all SNR values) slightly up to some $K_t$. This is expected because $\lim_{K_t \geq K_{to}} \mathrm{E}\{\chi^L_{max}(K_t)\} \approx c, \exists K_{to}$ for fixed $L$. Next we evaluate the effect of the number of RF chains on the performances of these approaches when $K_t = 32$ as shown in Fig. 6. From this figure, one can observe that increasing $N_a$ increases the average sum rate. Finally, we examine the effect of the number of transmitter antennas when $K_t = 32$ as shown in Fig, 7. From this figure, we also observe that increasing $N$ increases the average sum rate of the proposed HB approach which is in line with the theoretical result. However, the average sum rate of the existing ASB approach does not increase with $N$. This is due to the fact that the existing ASB approach employs only the first $N_a$ antennas. From Figs. 3 - 7, one can notice that the proposed HB approach achieves better performance than that of the existing ASB approach.

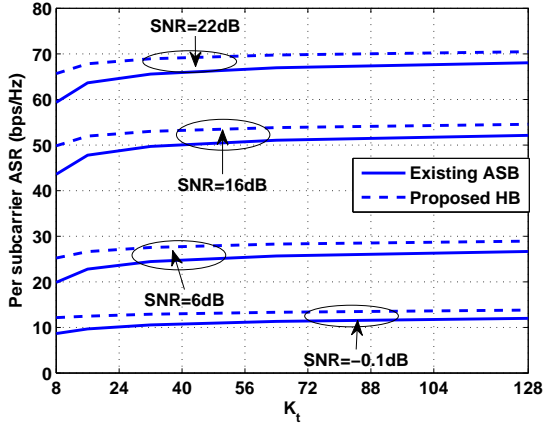 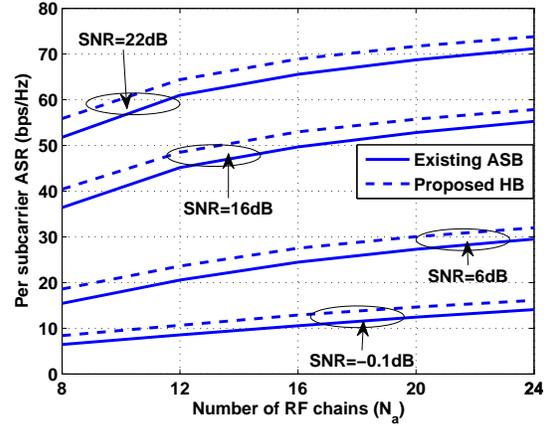

Fig. 5: The ASR of existing ASB and proposed HB approaches for different $K_t$.

Fig. 6: The ASR of existing ASB and proposed HB approaches for different $N_a$.

### B. Uniform Linear Array Channel

This subsection provides simulation results for the ULA channel model. To this end, we set $L_s = 8$, $K_i \leq K_{max}$, $K_t = 32$ and $N = 64$. Under such settings, we plot the sum rates obtained by existing ASB, proposed HB, and DB approaches for the following two cases.

**Case I**: In this case, we examine the average rates when $\theta_{km}, \forall m, k$ are taken randomly from a uniform distribution $\mathcal{U}[0, 2\pi]$ as shown in Fig. 8. As we can see from this figure, the proposed HB approach achieves significantly better performance than that of the existing ASB approach and superior performance is achieved by the DB approach which is expected.

**Case II**: For this case, we examine the sum rates of the aforementioned three approaches when $\theta_{km}, \forall m, k$ are selected as in the conditions stated by *Lemma 2* (Fig. 9). As we can see from this figure, the proposed HB approach achieves the same performance as that of DB and inferior performance is achieved by the existing ASB approach which is in line with *Lemma 2*.

The effects of $N$ and $N_a$ on the performances of the existing ASB and proposed HB for ULA channels can be studied like in the above subsection. The details are omitted for conciseness.

### C. Effect of the Number of Phase Shifters

Up to now, we employ the number of PSs as derived in Section IV-A. However, as motivated previously, it is practically interesting to realize the proposed HB architecture with limited number of CPPSs as in Section IV-B. This simulation examines the sum rate of the proposed HB for

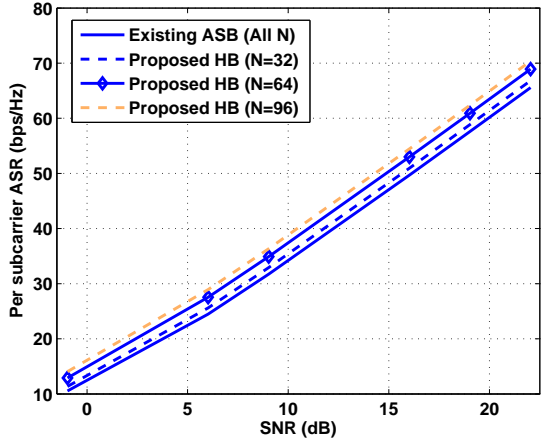 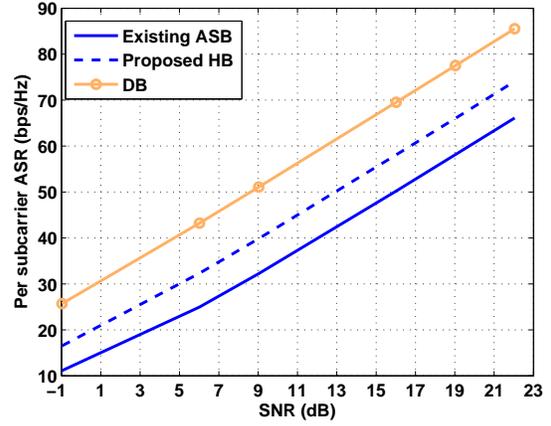

Fig. 7: The ASR of existing ASB and proposed HB approaches for different $N$.

Fig. 8: The ASR of existing ASB, proposed HB, and DB for ULA channels with AOD are taken from $\mathcal{U}[0,\ 2\pi]$.

$N = 128$ for different number of CPPSs per each RF chain (i.e., different levels of $\epsilon$) as shown in Fig. 10. As can be seen from this figure, the average sum rate saturates after a certain number of CPPSs which is around 40 for our setup. This demonstrates that the proposed HB can be realized with quite small number of CPPSs (i.e., from 20 to 40 CPPSs per RF chain) and hence it is suitable for practical implementation. When the number of CPPSs are zero, the proposed HB yields the same average sum rate as that of the existing ASB which is expected.

*D. Comparison of the Proposed and Existing Approaches for Flat fading Channel*

As detailed in the introduction section, a number of HB approaches are proposed where most of them employ MP algorithms. This motivates us to compare the performances of the proposed approach with those of [17] and [29]. The work of [17] proposes a HB for single user massive MIMO system with flat fading channel. The algorithm of this paper can be extended easily for multiuser setup when each receiver has single antenna by utilizing appropriate DB. Also in [29], a HB algorithm is proposed for flat fading multiuser massive MIMO setup. This simulation compares the algorithms of these papers, the existing ASB and the proposed HB algorithms. To this end, we take $N = 64$, $\rho_k = 1$, $K = 16$ (i.e., the number of served users) and employ ZF precoder for all algorithms (i.e., the proposed algorithm, and those of [17] and [29]). Fig. 11 shows the performances of these algorithms for ULA channel with different number of scatterers

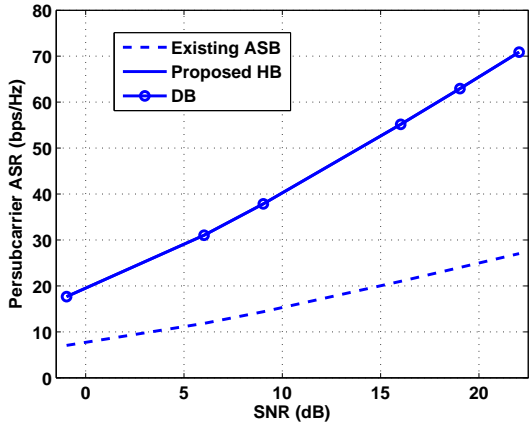
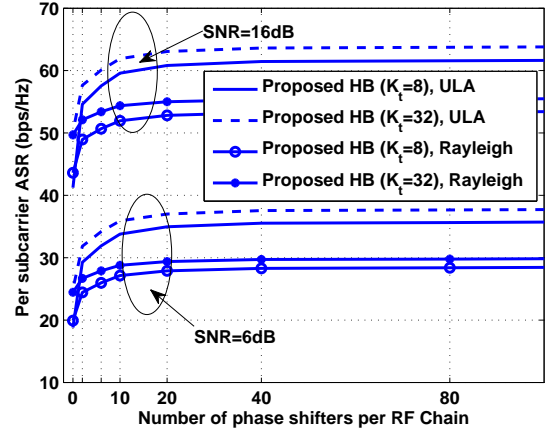

Fig. 9: The ASR of ASB, proposed HB, and DB for ULA channels with AODs are as in *Lemma 2* with $\theta = \frac{\pi}{2}$.

Fig. 10: The ASR of the proposed HB with different number of CPPSs per RF chain (Asymmetric signal flow).

$L_s$ and RF chains $N_a$. As can be seen from this figure, the performances of [29] and [17] are better than that of the ASB algorithm. However, the sum rates achieved by the algorithms of [29] and [17] are significantly lower than that of the DB especially when $L_s$ is large. For the asymmetric (Asy) signal flow case, the proposed HB achieves the same performance as that of the DB for both $N_a = 16$ and $24$ when $N_{PS} \geq 32$ (i.e., less than $N_{PS}$ of [29] and [17]). This figure also confirms that deploying only $N_{PS} = 16$ CPPSs per RF chain is still sufficient both for Asy and symmetric (Sym) signals flows. Hence, the proposed HB design is also cost efficient. Note that the algorithms presented in Fig. 11 have almost the same computational complexity.

## VIII. CONCLUSIONS

This paper considers hybrid beamforming for downlink multiuser massive MIMO systems in frequency selective channels. We examine the scenario where the transmitter equipped with $N$ antennas is serving $K$ decentralized single antenna users. For this scenario, first we quantify the required number of RF chains and PSs such that the proposed HB achieves the same performance as that of the DB which utilizes $N$ RF chains. We show that the performance obtained by the DB can be achieved with our HB just by utilizing $r_t$ RF chains and $2r_t(N - r_t + 1)$ PSs, where $r_t \leq N$ is the rank of the combined digital precoder matrices of all sub-carriers. Second, we provide simple and novel approach to reduce the number of PSs with negligible performance degradation. From simulation, we have found that only $20 - 40$ PSs per RF chain are sufficient for most practical parameter settings. Finally, for the case where the deployed number of RF chains $N_a < r_t$, we propose a simple user scheduling and sub-carrier allocation algorithm to choose the best set of served users in a sub-carrier. The performance of the proposed scheduling algorithm is examined analytically. Extensive numerical simulations are performed to validate theoretical results, and study the effects of different parameters such as $N_a$, $N$ and PSs. Computer simulations also demonstrate that the proposed HB achieves significantly better performance than those of the existing HBs in both flat fading and frequency selective channels. Moreover, our HB design is simple and convenient for practical implementation of massive MIMO systems.

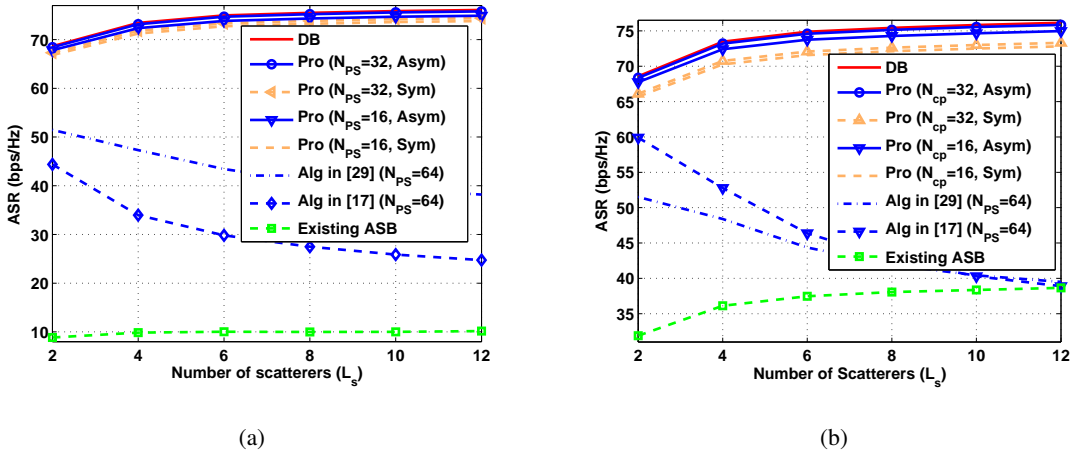

Fig. 11: Comparison of the proposed and existing algorithms for flat fading ULA channel at $SNR = 10$dB: (a) when $N_a = 16$, (b) when $N_a = 24$. In this figure, $N_{PS}$ denotes the number of PSs per RF chain, $SNR = \frac{P}{\sigma^2}$ and $\sigma^2 = 1$mW.

## APPENDIX A: PROOF OF *Lemma 1*

For convenience, we provide the proof of *Lemma 1* by omitting the superscript $(.)^d$ in $\mathbf{B}_i^d$

*Existing ASB approach*

When the beamforming matrix of each sub-carrier $\mathbf{B}_i$ has $N_a$ rows, the rank constraint of (15) is satisfied implicitly. Under this setting, the user scheduling can be performed per sub-carrier independently. The remaining task is to examine this problem for each sub-carrier.

$$\max_{\mathbf{B}_i^{ASB}, p_{ik}^{ASB}} \sum_{k=1}^{K_i} \log(1 + p_{ik}^{ASB}), \text{ s.t } \mathbf{H}_i^H(\mathbb{K}_i^{ASB})\mathbf{B}_i^{ASB} = \mathbf{I}, \sum_{k=1}^{K_i} p_{ik}^{ASB}[(\mathbf{B}_i^{ASB})^H \mathbf{B}_i^{ASB}]_{k,k} \leq P_i \quad (19)$$

where $\mathbf{H}_i(\mathbb{K}_i^{ASB}) \in \mathcal{C}^{N_a \times K_i}$ is the truncated channel matrix of the users of sub-carrier $i$ scheduled by the existing ASB. As the total number of users $K_t$ is very large, at optimality $\text{rank}(\mathbf{H}_i(\mathbb{K}_i^{ASB})) = N_a$ is satisfied almost surely. Thus, without loss of generality, we assume that $\mathbf{H}_i(\mathbb{K}_i^{ASB})$ is a full rank channel matrix. Under the ZF beamforming design, we have

$$\mathbf{B}_i^{ASB} = \mathbf{H}_i(\mathbb{K}_i^{ASB})[\mathbf{H}_i(\mathbb{K}_i^{ASB})^H \mathbf{H}_i(\mathbb{K}_i^{ASB})]^{-1}. \quad (20)$$

By employing $\mathbf{B}_i^{ASB}$ and performing some mathematical manipulations, the power allocation part of (19) can be re-expressed as

$$R_i^{ASB} = \max_{p_{ik}^{ASB}, \forall i,k} \sum_{k=1}^{K_i} \log_2(1 + p_{ik}^{ASB} g_{ik}^{ASB}), \text{ s.t} \sum_{k=1}^{K_i} p_{ik}^{ASB} = P_i \quad (21)$$

where $\mathbf{b}_{ik}^{ASB}$ is the $k$th column of $\mathbf{B}_i^{ASB}$ and

$$g_{ik}^{ASB} = \frac{1}{|\mathbf{b}_{ik}^{ASB}|^2}. \quad (22)$$

From ZF precoding and $\text{rank}(\mathbf{B}_i^{ASB}) = K_i$, the $k$th column of $\mathbf{B}_i^{ASB}$ ($\mathbf{b}_{ik}^{ASB}$) satisfies

$$\mathbf{h}_{im}(\mathbb{K}_i^{ASB})^H \mathbf{b}_{ik}^{ASB} = \delta_{k,m}, \ m = 1, \cdots, K_i \quad (23)$$

where $\delta_{k,m}$ is the Dirac delta and $\mathbf{h}_{im}(\mathbb{K}_i^{ASB})$ is the $m$th column of $\mathbf{H}_i(\mathbb{K}_i^{ASB})$. Thus, $\mathbf{b}_{ik}^{ASB}$ should be orthogonal to the sub-space $\mathcal{B}_{ik}^{ASB} = \text{span}\{\mathbf{h}_{im}(\mathbb{K}_i^{ASB}) : m = 1, \cdots, K_i, m \neq k\}$. It follows

$$\mathbf{b}_{ik}^{ASB} = \frac{\mathbf{h}_{ik}^H(\mathbb{K}_i^{ASB})\boldsymbol{\Gamma}_{ik}^{\perp}(\mathbb{K}_i^{ASB})}{\mathbf{h}_{ik}^H(\mathbb{K}_i^{ASB})\boldsymbol{\Gamma}_{ik}^{\perp}(\mathbb{K}_i^{ASB})\mathbf{h}_{ik}(\mathbb{K}_i^{ASB})}, \ g_{ik}^{ASB} = \frac{1}{|\mathbf{b}_{ik}^{ASB}|^2} = |\mathbf{h}_{ik}^H(\mathbb{K}_i^{ASB})\boldsymbol{\Gamma}_{ik}^{\perp}(\mathbb{K}_i^{ASB})|^2 \quad (24)$$

where $\boldsymbol{\Gamma}_{ik}^{\perp}(\mathbb{K}_i^{ASB})$ is the orthogonal projector of $\mathcal{B}_{ik}^{ASB}$ and the third equality holds due to the fact that any orthogonal projector is idempotent [52].

*Proposed HB approach*

In the proposed approach, the combined precoder matrix $\mathbf{B}^{HB}$ obtained by **Algorithm I** will have a rank of $N_a$. By applying similar technique as above, the rate achieved by the proposed approach can be obtained by solving the following optimization problem

$$R_i^{HB} = \max_{p_{ik}^{HB}, \forall i,k} \sum_{k=1}^{K_i} \log_2(1 + p_{ik}^{HB} g_{ik}^{HB}), \quad \text{s.t} \sum_{k=1}^{K_i} p_{ik}^{HB} = P_i \tag{25}$$

where

$$\mathbf{b}_{ik}^{HB} = \frac{\mathbf{h}_{ik}^H(\mathbb{K}_i^{HB})\mathbf{\Gamma}_{ik}^\perp(\mathbb{K}_i^{HB})}{\mathbf{h}_{ik}^H(\mathbb{K}_i^{HB})\mathbf{\Gamma}_{ik}^\perp(\mathbb{K}_i^{HB})\mathbf{h}_{ik}(\mathbb{K}_i^{HB})}, \quad g_{ik}^{HB} = \frac{1}{|\mathbf{h}_{ik}^H(\mathbb{K}_i^{HB})\mathbf{\Gamma}_{ik}^\perp(\mathbb{K}_i^{HB})|^2}, \tag{26}$$

$\mathbf{\Gamma}_{ik}^\perp(\mathbb{K}_i^{HB})$ is the orthogonal projector of $\mathcal{B}_{ik}^{HB} = \text{span}\{\mathbf{h}_{im}(\mathbb{K}_i^{HB}) : m = 1, 2, \cdots, K_i, m \neq k\}$ and $\mathbf{b}_{ik}^{HB}$ is the $k$th column of $\mathbf{B}_i^{HB}$.

In both the proposed HB and existing ASB approaches, the number of served users of subcarrier $i$ is $K_i$. Furthermore, when $\mathbb{K}_i^{HB} = \mathbb{K}_i^{ASB}, \forall i$, $(\mathcal{B}_{ik}^{HB})^\perp$ is a superset of $(\mathcal{B}_{ik}^{ASB})^\perp$ (i.e., $(\mathcal{B}_{ik}^{HB})^\perp \supset (\mathcal{B}_{ik}^{ASB})^\perp, \forall i$). This is due to the fact that the dimension of $\mathbf{h}_{im}(\mathcal{B}_{ik}^{HB})^\perp$ ($N \times 1$) is larger than that of $\mathbf{h}_{im}(\mathcal{B}_{ik}^{ASB})^\perp$ which is $N_a \times 1$. For these reasons, we will have

$$g_{ik}^{HB} \geq g_{ik}^{ASB} \Rightarrow R_i^{HB} \geq R_i^{ASB}, \quad \text{when } \mathbb{K}_i^{HB} = \mathbb{K}_i^{ASB}.$$

APPENDIX B: PROOF OF *Theorem 2*

When $\mathbb{K}_i^{ASB} \neq \mathbb{K}_i^{HB} \neq \mathbb{K}_i^{DB}$ for some $i$, the relation between $R_i^{ASB}$, $R_i^{HB}$ and $R_i^{Di}$ cannot be quantified for each channel realization. Thus, we compare these three approaches by examining their average sum rates by assuming ZF precoding and equal power allocation strategy. For better exposition of the proof of *Theorem 2*, let us consider the following *Lemma*.

*Lemma C.1*: Let $\mathbf{A} = \mathbf{Z}^H\mathbf{Z}$ be a non singular hermitian matrix of size $K \times K$ and $\mathbf{B} = \mathbf{A}^{-1}$, where $\mathbf{Z} \in \mathcal{C}^{N_a \times K}$ and $K \leq N_a$. We partition $\mathbf{A}$ and $\mathbf{B}$ as

$$\mathbf{A} = \begin{bmatrix} a_{11} & \mathbf{a}_{21}^H \\ \mathbf{a}_{21} & \mathbf{A}_{22} \end{bmatrix}, \quad \mathbf{B} = \begin{bmatrix} b_{11} & \mathbf{b}_{12} \\ \mathbf{b}_{21} & \mathbf{B}_{22} \end{bmatrix} \tag{27}$$

where $a_{11}(b_{11})$ is a scalar value, and the rest of the terms are appropriate dimension vectors or matrices. If $a_{11} \neq 0$ and $\mathbf{A}_{22}$ is non singular, we can express $b_{11}$ as

$$\frac{1}{b_{11}} = a_{11} - \mathbf{a}_{21}^H \mathbf{A}_{22}^{-1} \mathbf{a}_{21}. \tag{28}$$

And, if each element of $\mathbf{Z}$ is taken from i.i.d ZMCSCG random variable with variance 1, then

$$\frac{1}{b_{11}} \sim \chi^2_{N_a-K+1} \tag{29}$$

*Proof:* The first equality (28) can be proved by applying the well known Schur Complement theorem. The detailed derivation can also be found from *Theorem A5.2* of [53].

To prove (29) we note that both $a_{11}$ and $\mathbf{a}_{21}^H \mathbf{A}_{22}^{-1} \mathbf{a}_{21}$ are strictly non negative real values. And when $a_{11} - \mathbf{a}_{21}^H \mathbf{A}_{22}^{-1} \mathbf{a}_{21}$ is a non negative real valued term, by applying *Theorem 3.2.10* of [53], the probability density function of $\frac{1}{b_{11}}$ can be expressed as $\mathcal{W}_1(N_a - K + 1, 1)$, where $\mathcal{W}_{(.)}(.,.)$ denotes a real valued Wishart distribution. It follows

$$\frac{1}{b_{11}} \sim \mathcal{W}_1(N_a - K + 1, 1) \sim \chi^2_{N_a-K+1} \tag{30}$$

where the second distribution is due to the fact that $\mathcal{W}_1(N_a - K + 1, 1)$ has the same distribution as that of Chi-square ($\chi^2$) distribution with $N_a - K + 1$ degrees of freedom (see *Corollary 3.2.2* of [53]). As expected when $N_a = K$, $\frac{1}{b_{11}}$ is a $\chi^2$ distribution with 1 degree of feedom. ∎

In the following we prove (18). By setting $\mathbf{Z}$ of *Lemma 3* as $\mathbf{Z} = \mathbf{H}_i(\mathbb{K}_i^{ASB})$, we get $\frac{1}{[(\mathbf{H}_i(\mathbb{K}_i^{ASB})^H \mathbf{H}_i(\mathbb{K}_i^{ASB}))^{-1}]_{1,1}} \sim \chi^2_{N_a-K+1}$. Hence

$$x_i^{ASB} \triangleq \frac{1}{[(\mathbf{H}_i(\mathbb{K}_i^{ASB})^H \mathbf{H}_i(\mathbb{K}_i^{ASB}))^{-1}]_{k,k}} \sim \chi^2_{N_a-K+1}.$$

Since $\log(1+x)$ is a concave function, by employing Jensen's inequality, we will have

$$\mathrm{E}\{R_{ik}^{ASB}\} \leq \log_2\left(1 + \frac{P}{K}\mathrm{E}\{x_i^{ASB}\}\right) \tag{31}$$

The current paper employs scheduling of $K_t \geq K_i$ users and when we have $K_t$ users, there are $K_g = \lceil \frac{K_t}{K} \rceil$ independent groups. And the proposed approach selects a group having maximum sum rate which is directly related to $x_i^{ASB}$. Thus, a group will achieve the best maximum sum rate if its $x_i^{ASB}$ is the highest of all of these $K_g$ groups. Therefore, $\mathrm{E}\{R_{ik}^{ASB}\}$ is bounded as

$$\mathrm{E}\{R_{ik}^{ASB}\} \leq \log_2\left(1 + \frac{P}{K}\mathrm{E}\{x_{i_{max}}^{ASB}(K_g)\}\right) \tag{32}$$

where $x_{i_{max}}^{ASB}(K_g) = \max\{x_i^{ASB}(1), x_i^{ASB}(1), \cdots, x_i^{ASB}(K_g)\}$ is the maximum of $K_g$ independent Chi-square distributed random variable with $N_a - K + 1$ degrees of freedom.

In the following, we valuate $\mathrm{E}\{x_{i_{max}}^{ASB}(K_g)\}$. By applying order statistics, the probability density function (pdf) of $z_{max} \triangleq x_{i_{max}}^{ASB}(K_g)$ can be expressed as [54]

$$f_{z_{max}}(x) = K_g(F(x))^{K_g-1}f(x) \tag{33}$$

where

$$F(x) = \frac{\gamma(\frac{N_a-K+1}{2}, \frac{x}{2})}{\mathbf{\Gamma}(\frac{N_a-K+1}{2})}, \quad f(x) = \frac{1}{2^{\frac{N_a-K+1}{2}}\mathbf{\Gamma}(\frac{N_a-K+1}{2})}\mathbf{x}^{\frac{N_a-K+1}{2}-1}e^{\frac{-x}{2}} \quad (34)$$

with $\mathbf{\Gamma}(.)$ is the Gamma function and $\gamma(.)$ as the lower incomplete Gamma function. It follows

$$\mathrm{E}\{x_{i_{max}}^{ASB}(K_g)\} = K_g \int_0^\infty (F(x))^{K_g-1} f(x) dx. \quad (35)$$

When $K_g = 1$, $\mathrm{E}\{x_{i_{max}}^{ASB}(K_g)\} = N_a - K + 1$. However, for general $K_g$, getting closed form solution for this integral is non trivial. Due to this reason, we utilize numerical approach to evaluate this integral (for example simple trapezoid numerical integration approach of Matlab).

Like in (32), one can also get the following upper bound rate for the DB approach

$$\mathrm{E}\{R_{ik}^{DB}\} \leq \log_2\left(1 + \frac{P}{K}\mathrm{E}\{x_{i_{max}}^{DB}(K_g)\}\right) \quad (36)$$

where $\mathrm{E}\{x_{i_{max}}^{DB}(K_g)\}$ is the expected value of the maximum of $K_g$ independent $\chi^2_{N-K-1}$ random variables. And for the proposed HB approach, we will have the following rates.

$$\mathrm{E}\{R_{ik}^{HB}\} \leq \log_2(1 + \frac{P}{K}\mathrm{E}\{x_{i_{max}}^{HB_1}(K_s)\}) \; i \leq \tilde{S}, \quad \mathrm{E}\{R_{ik}^{HB}\} \leq \log_2(1 + \frac{P}{K}\mathrm{E}\{x_{i_{max}}^{HB_2}(K_g)\}) \; i > \tilde{S} \quad (37)$$

where $K_s = \lceil \frac{N_f K_t}{N_a K} \rceil$ and $\mathrm{E}\{x_{i_{max}}^{HB_1}(K_s)\}$ ($\mathrm{E}\{x_{i_{max}}^{HB_2}(K_g)\}$) is the expected value of the maximum of $K_s(K_g)$ independent $\chi^2_{N-K+1}$ ($\chi^2_{N_a-K+1}$) random variables. By substituting (32), (36) and (37) into the average sum rate expressions of all sub-carriers, we get (18).

APPENDIX C: PROOF OF *Lemma 3*

In the following, we provide channel matrices that satisfy $\mathrm{rank}(\mathbf{H} = [\mathbf{H}_1, \mathbf{H}_2, \cdots, \mathbf{H}_{N_f}]) \lessapprox N_a$ for the ULA multipath channel models. By employing the multipath and ULA channel models (4) and (5), and after doing some mathematical manipulations, $\mathbf{h}_{ik}$ can be expressed as

$$\mathbf{h}_{ik} = \boldsymbol{\tau}_k \mathbf{C}_k \tilde{\mathbf{f}}_i \quad (38)$$

where $\mathbf{C}_k = [\mathbf{c}_k(0), \mathbf{c}_k(1), \cdots, \mathbf{c}_k(L_p-1)]$, $\tilde{\mathbf{f}}_i = [1, e^{j\frac{2\pi i}{N_f}}, e^{j\frac{2\pi i}{N_f}2}, \cdots, e^{j\frac{2\pi i}{N_f}(L_p-1)}]^T$, $j = \sqrt{-1}$ and $\boldsymbol{\tau}_k$ is as defined in (4). Using (38), one can rewrite $\mathbf{H} = \boldsymbol{\tau}\mathbf{C}\tilde{\mathbf{F}}$, where $\boldsymbol{\tau} = [\boldsymbol{\tau}_1, \boldsymbol{\tau}_2, \cdots, \boldsymbol{\tau}_{K_t}]$, $\mathbf{C} = \mathrm{blkdiag}(\mathbf{C}_1, \mathbf{C}_2, \cdots, \mathbf{C}_{K_t})$ and $\tilde{\mathbf{F}} = [\tilde{\mathbf{F}}_1, \tilde{\mathbf{F}}_2, \cdots, \tilde{\mathbf{F}}_{N_f}]$ with $\tilde{\mathbf{F}}_i = \mathbf{I}_{K_t} \otimes \tilde{\mathbf{f}}_i$. For any $\mathbf{C}$ and $\tilde{\mathbf{F}}$, since $\mathrm{rank}(\mathbf{H}) \leq \mathrm{rank}(\boldsymbol{\tau})$, one can maintain $\mathrm{rank}(\mathbf{H}) \leq N_a$ just by ensuring $\mathrm{rank}(\boldsymbol{\tau}) \lessapprox N_a$.

As can be seen from (4), for the given $\theta_{km}, m = 1, \cdots, L_s$, $\tilde{\boldsymbol{\tau}}_k(\theta_{km})$ is a Fourier vector with resolution $\frac{1}{N}$. Hence, when $L_p \leq N_a$, $L_s \leq N_a$, $\tilde{d} = \frac{\lambda}{2}$ and the AOD of the $K_t$ users satisfy $\sin(\theta_{km}) \in n\sin(\theta)[-\frac{1}{2N}, \frac{1}{2N}]$, $n = 1, 2, \cdots, N_a$, each column of $\boldsymbol{\tau}$ will be a Fourier vector with bounded $\sin(\theta_{km}) \in [-\frac{1}{2N}\sin(\theta), (N_a + \frac{1}{2N})\sin(\theta)]$. Thus, the columns of $\boldsymbol{\tau}$ will contain a maximum of $N_a$ orthogonal (i.e., linearly independent) Fourier vectors. Consequently, each column of $\boldsymbol{\tau}$ can be well approximated by linearly combining $N_a$ orthogonal Fourier vectors in the practically relevant $N_a < N$. Specifically, as $N \to \infty$ and fixed $N_a$, $\sin(\theta_{km}) \in n\sin(\theta), n = 1, 2, \cdots, N_a$ and the columns of $\boldsymbol{\tau}$ will lie exactly to a maximum of $N_a$ orthogonal Fourier vectors. For these reasons, we will have $\text{rank}(\boldsymbol{\tau}) \lessapprox N_a$ (and $\text{rank}(\boldsymbol{\tau}) \leq N_a$ when $N \to \infty$).

## IX. MORE DERIVATIONS AND RESULTS

### A. Derivation of (8)

For convenience, let us again recompute the QR decomposition of $(\mathbf{Q}^d)^H$ as

$$(\mathbf{Q}^d)^H = \bar{\mathbf{B}}[\bar{\bar{\mathbf{B}}} \ \tilde{\bar{\mathbf{B}}}] = \bar{\mathbf{B}}\bar{\bar{\mathbf{B}}}\boldsymbol{\alpha}[\boldsymbol{\alpha}^{-1} \ \boldsymbol{\alpha}^{-1}\bar{\bar{\mathbf{B}}}^{-1}\tilde{\bar{\mathbf{B}}}] = \tilde{\mathbf{B}}^H\tilde{\mathbf{A}}^H \tag{39}$$

where $\tilde{\mathbf{A}} \in \mathbb{C}^{N \times r_t}$, $\tilde{\bar{\mathbf{B}}} \in \mathbb{C}^{r_t \times r_t}$ is an upper triangular full rank matrix, $\tilde{\mathbf{B}} = (\bar{\mathbf{B}}\bar{\bar{\mathbf{B}}}^{-1}\boldsymbol{\alpha})^H$, $\tilde{\mathbf{A}} = [\boldsymbol{\alpha}^{-1} \ \boldsymbol{\alpha}^{-1}\bar{\bar{\mathbf{B}}}^{-1}\tilde{\bar{\mathbf{B}}}]^H$ and $\boldsymbol{\alpha}$ is an introduced diagonal scaling factor matrix ensuring that each entry of $\tilde{\mathbf{A}}$ has a maximum amplitude of 2. By substituting (39) into (7), we will have (8).

### B. Closed form expression for $\text{E}\{\chi^N_{max}(2)\}$ in (18)

Let $x_1, x_2, \cdots x_N$ are $N$ i.i.d random variables. The CDF of $x_{max} = \max\{x_1, x_2, \cdots, x_N\}$ is

$$F_{max}(x_1, x_2 \cdots, x_N)(z) = P(x_{max} \leq z) = P(x_1 \leq z, x_2 \leq z, \cdots, x_N \leq z) = F_x(z)^N.$$

Furthermore, the pdf of $x_{max}$ is given as

$$f_{max}(z) = \frac{dF_{max}}{dz} = NF_x(z)^{N-1}\frac{dF_x(z)}{dz} = NF_x(z)^{N-1}f_x(z)$$

where $f_x(z) = \frac{dF_x(z)}{dz}$. It follows that

$$\text{E}\{f_{max}\} = \int zf_{max}(z)dz.$$

Now if $x_i, \forall i$ are Chi-square distributed random variables each with 2 degrees of freedom, we will have

$$F(z) = 1 - e^{-z/2}, \quad f(z) = \frac{1}{2}e^{-z/2}.$$

By employing binomial expansion, one can get

$$F(z)^{(N-1)} = [1 - e^{-z/2}]^{(N-1)} = \sum_{k=0}^{N-1}(-1)^k \binom{N-1}{k} e^{-kz/2}.$$

It follows that

$$\mathrm{E}\{f_{max}\} = \frac{1}{2}\sum_{k=0}^{N-1}(-1)^k \binom{N-1}{k} \int ze^{-\frac{(k+1)z}{2}}dz$$

It follows

$$\int ze^{-\frac{(k+1)z}{2}}dz = \int \frac{t}{a^2}e^{-t}dt = \frac{1}{a^2}\int_0^\infty te^{-t}dt = \frac{1}{a^2}\Gamma(2) = \frac{1}{a^2}$$

where $a = \frac{(k+1)}{2}$, $t = az$ and $\Gamma(.)$ is a Gamma function. Thus

$$\mathrm{E}\{f_{max}\} = 2\sum_{k=0}^{N-1}(-1)^k \binom{N-1}{k} \frac{1}{(k+1)^2}. \tag{40}$$

When $N = 1$, $\mathrm{E}\{f_{max}\} = 2$ which is expected.

## C. Simulation Results for realistic channel model for ULA and UPA antenna arrays

This section provides simulation results for the realistic channel model obtained via extensive experiments at 28 and 73 GHz carrier frequencies in [45]. In the current paper, we consider the model presented in (9) of [45] while ignoring the Doppler shift. In this regard, we consider two antenna array structures one is the ULA (as presented in the revised manuscript) and the other is the uniform planar antenna array (UPA) as defined in (7) of [17]. Under such settings, an Urban Micro BS (UMi) single cell model having a hexagonal cell structure with cell radius 100 m is considered. Like in [45], we assume that each user equipment (UE) is distributed uniformly inside the radius of the cell. All the other system parameters are summarized as shown in Table I below.

Fig. 12 - Fig. 15 shows the rate achieved by the DB, proposed HB and the algorithms of [17], [29] for single user and multiuser systems. As can be seen from these figures, the approaches of [17], [29] achieve almost the same performance as that of the DB when $K = 1$ (i.e., single user case) and UPA antenna array models, and there is a slight decrease in performance for the ULA antenna array models. However, in the multiuser setup, the approaches of these papers achieve poor performance. One of the potential reasons for this could be in a multiuser setup,

TABLE I: Simulation parameters for the practical channel models.

| Radius of the cell | 100 m |
|---|---|
| Number of antennas at the BS | 64 (ULA 64 × 1) or (UPA 8 × 8) |
| Number of antennas at each UE | 1 |
| Carrier frequency | 28 GHz |
| Micro BS transmitter power ($P_{BS}$) | {50,40,30,20,10,0}dBm |
| System bandwidth | 500 MHz |
| Noise figure at each user | 7 dB |
| Receiver temperature | 300 K |

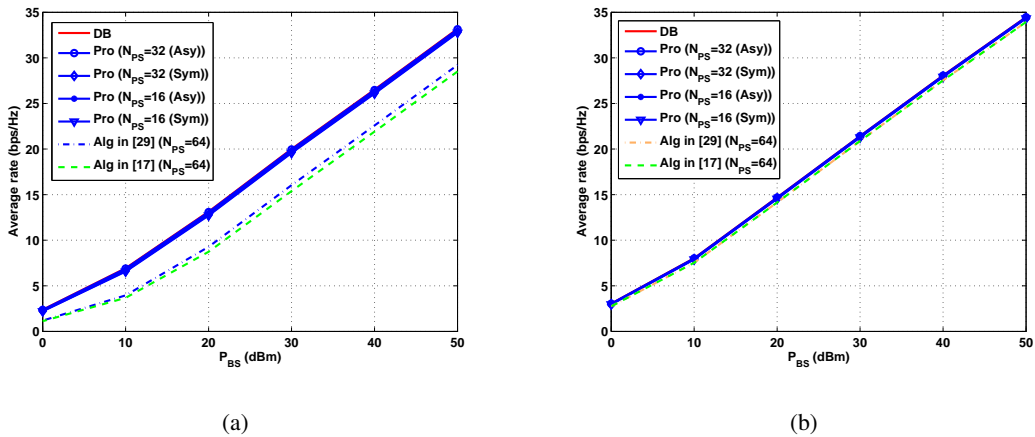

(a)

(b)

Fig. 12: Comparison of the proposed and existing algorithms for flat fading channel when $N_a = 1$ and $K = 1$ (i.e., single user case): (a) For ULA antenna array $64 \times 1$, (b) For UPA antenna array $8 \times 8$.

two or more UEs may align in the same direction. Consequently, the beam pattern that is aligned to the desired UE will create strong interference to other UEs in the same/similar directions. Nevertheless, the proposed approach achieves almost the same performance as that of the DB for all settings with asymmetric (Asy) and symmetric (Sym) signal flow cases under the ULA and UPA antenna array models. Furthermore, this simulation also demonstrates that the proposed approach utilizes small number of PSs compared to those of [17], [29].

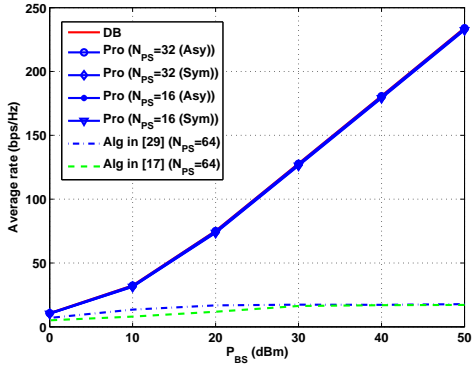
(a)

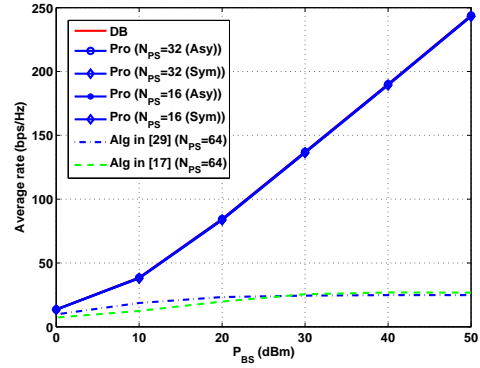
(b)

Fig. 13: Comparison of the proposed and existing algorithms for flat fading channel when $N_a = 8$ and $K = 8$ (i.e., multiuser case): (a) For ULA antenna array $64 \times 1$, (b) For UPA antenna array $8 \times 8$.

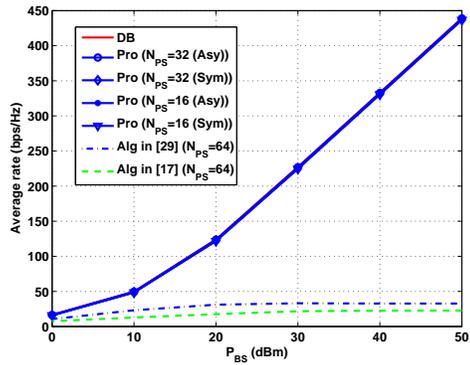
(a)

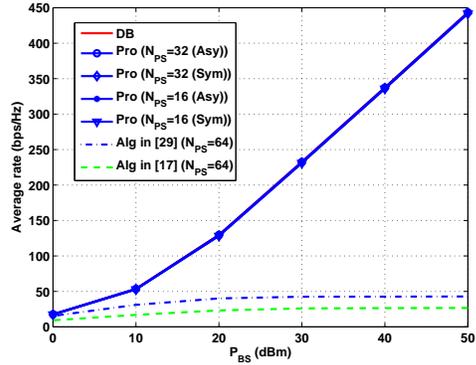
(b)

Fig. 14: Comparison of the proposed and existing algorithms for flat fading channel when $N_a = 16$ and $K = 16$ (i.e., multiuser case): (a) For ULA antenna array $64 \times 1$, (b) For UPA antenna array $8 \times 8$.

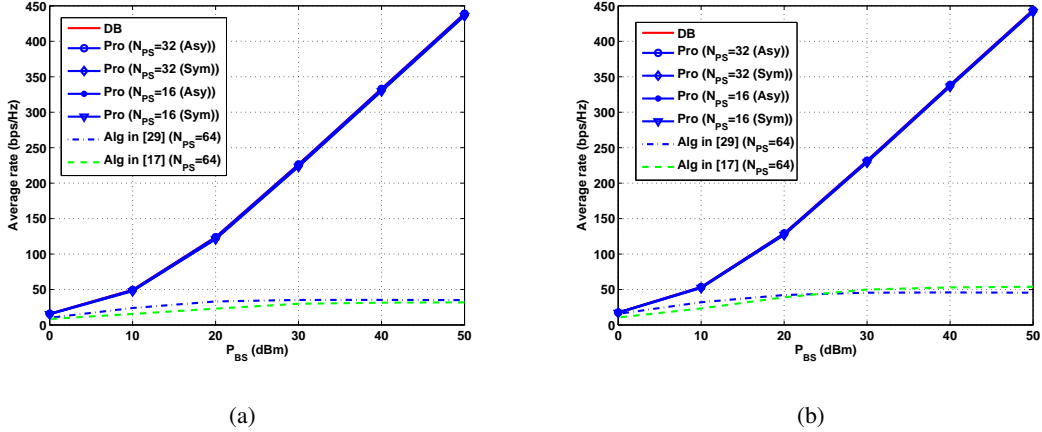

Fig. 15: Comparison of the proposed and existing algorithms for flat fading channel when $N_a = 24$ and $K = 16$ (i.e., multiuser case): (a) For ULA antenna array $64 \times 1$, (b) For UPA antenna array $8 \times 8$.